\documentclass[journal=jpcbfk, manuscript=article, layout=twocolumn]{achemso}

\usepackage{graphicx}
\usepackage[version=4]{mhchem}
\usepackage{dcolumn}
\usepackage{bm}
\usepackage{multirow}
\usepackage{url}
\usepackage{tablefootnote}
\usepackage{etoolbox}

\usepackage[hidelinks]{hyperref}
\usepackage{siunitx}

\setkeys{acs}{doi=true}

\usepackage{color}
\usepackage{amsmath}

\title{Distinguishing sodium-ion penetration and sustained transport in realistic hard carbon nanostructures}
 
\author{Carolina Cruz-Cardona}
\altaffiliation{These authors contributed equally to this work.}
\affiliation{Department of Engineering, King's College London, London, United Kingdom}

\author{Andreea Hedes}
\altaffiliation{These authors contributed equally to this work.}
\affiliation{ University of Manchester, United Kingdom}

\author{Patrick Rowe}
\affiliation{SandboxAQ, Palo Alto, CA, United States}

\author{Anderson Arboleda-Lamus}
\affiliation{Department of Physics, King's College London, London, United Kingdom}

\author{Raul Arenal}
\affiliation{Instituto de Nanociencia y Materiales de Aragon, University of Zaragoza-CSIC, Spain}

\author{Zhenyu Guo}
\affiliation{Department of Chemical Engineering, Imperial College London, London, United Kingdom}

\author{Heather Au}
\affiliation{Department of Chemical Engineering, Imperial College London, London, United Kingdom}

\author{Magda M. Titirici}
\affiliation{Department of Chemical Engineering, Imperial College London, London, United Kingdom}

\author{Celine Merlet}
\affiliation{CIRIMAT, University of Toulouse, Toulouse INP, CNRS, France}

\author{Carla de Tomas}
\email{carla.de\_tomas@kcl.ac.uk}
\affiliation{Department of Physics, King's College London, London, United Kingdom}

\begin{document}

\begin{abstract}
 
Hard carbons are promising anodes for sodium-ion energy storage, yet the atomistic relationship between their heterogeneous microstructure and sodium transport remains poorly understood. Here, we combine machine-learning-generated hard-carbon structures with molecular dynamics simulations of explicit 1 M NaPF$_6$ in an ethylene carbonate/dimethyl carbonate electrolyte to investigate sodium storage and transport across carbon structures spanning densities from 0.5 to 2.0 g cm$^{-3}$. Increasing carbon density produces progressively less accessible pore networks and more graphitic local ordering, enabling the influence of microstructure on electrochemical response and Na$^+$ transport to be examined systematically. Sodium uptake increases with applied potential but decreases strongly with carbon density, while electrode charging persists even in the densest structure where Na$^+$ uptake remains very small. Analysis of Na–C and Na–O coordination shows increasing interaction with the carbon framework under applied potential, while Na$^+$ retains a partially solvated local environment. Three-dimensional trajectory analysis further reveals heterogeneous and predominantly tortuous transport pathways, with the more open HC-C1 structure supporting broader penetration and greater sustained displacement. Importantly, maximum penetration depth and retained displacement are not equivalent: ions can transiently reach deep regions of the denser carbon structures before returning towards the electrolyte interface. A one-dimensional potential-of-mean-force analysis further shows that deep penetration occurs across a range of projected free-energy changes and therefore cannot be described by a single energetic criterion. These results demonstrate that Na$^+$ storage and transport are governed by the interplay between electrostatic driving force, electrolyte solvation, pore accessibility and confinement. The findings highlight carbon microstructure as a key determinant of whether electrostatic polarisation translates into sustained ionic penetration and provide an atomistic framework for relating hard-carbon structure to sodium-storage behaviour.

\end{abstract}



\section{Introduction}

Sodium-ion energy storage technologies, including sodium-ion batteries (SIBs) and sodium-ion capacitors, have emerged as promising alternatives to lithium-ion systems for large-scale stationary energy storage owing to the natural abundance, low cost and widespread availability of sodium resources \cite{yabuuchi2014research,hwang2017sodium,cai2021comprehensive,vaalma2018cost}. Among the various anode materials investigated, hard carbons (HCs) remain the leading candidates because they combine high reversible capacities, excellent cycling stability and compatibility with commercially relevant electrolytes. Nevertheless, despite considerable experimental and computational progress, the atomistic relationship between hard carbon microstructure and sodium storage remains incompletely understood. This knowledge gap arises primarily from the intrinsically disordered and heterogeneous nature of these materials, whose complex microstructure makes it difficult to establish quantitative relationships between  local structural features, sodium uptake and  electrochemical behaviour \cite{Wu2024}.

Hard carbon is commonly described as a non-graphitising carbon composed of a heterogeneous arrangement of interconnected graphitic and porous nanodomains \cite{Stevens_2000,bommier2015new}. Rather than exhibiting a well-defined crystalline structure, hard carbons consist of curved and defective graphene fragments with varying degrees of local ordering, giving rise to a three-dimensional porous network containing both open and closed pores spanning a broad range of sizes and connectivities. Within these porous regions, locally slit-like arrangements may coexist with highly curved graphitic domains, further blurring the distinction between graphitic interlayers and nanopores. This structural diversity is compounded by the presence of vacancies, edge sites and surface functional groups, resulting in a broad distribution of adsorption environments and ion transport pathways. Consequently, sodium storage and transport cannot be understood without explicitly accounting for the heterogeneous multiscale architecture of hard carbons.

The influence of this structural heterogeneity is reflected in the characteristic voltage–capacity profile of hard carbon electrodes, which has traditionally been divided into a high-voltage sloping region and a low-voltage plateau. The sloping region has generally been associated with sodium adsorption at defects and disordered graphitic surfaces, whereas the plateau has been attributed either to sodium accumulation within nanopores or to interlayer insertion between stacked graphene sheets \cite{Stevens_2000,Komaba2011}. Increasingly, however, experimental and atomistic studies indicate that these descriptions are not mutually exclusive and that sodium storage involves a continuous interplay between adsorption, interlayer insertion and pore filling, with their relative contributions depending on the local carbon structure.\cite{Eren2025,Muck2025,Olsson2025,Front2026,KitsuIglesias2025} Recent computational studies have provided substantially more quantitative insight into these contributions. In particular, Gran Canonical Monte Carlo simulations have shown that the balance between adsorption, intercalation and pore filling varies systematically with carbon density and pore accessibility.\cite{Front2026} In a complementary study, Wang et al.\cite{Wang2026} combined a machine-learning potential with a random-forest site-identification approach to classify individual sodium environments and resolve their contributions to the voltage and capacity during progressive sodium insertion. These studies substantially advance the atomistic understanding of sodium-storage mechanisms and demonstrate that the structural environment of the carbon is a key determinant of sodium uptake.

Atomistic simulations have been instrumental also in demonstrating that defects act as preferential adsorption and nucleation sites for sodium, while pore size critically controls adsorption energetics, ion diffusion and cluster formation \cite{deTomas2023,Lin2025,Li2022b}. Nanopores with characteristic dimensions of approximately 1–2 nm have repeatedly been identified as particularly favourable because they simultaneously promote adsorption at pore surfaces and stabilise sodium clusters within confined volumes \cite{Lin2025}. Reactive molecular dynamics simulations additionally show that sodium insertion induces local structural rearrangements, including interlayer expansion and mechanical degradation, illustrating the intimate coupling between sodium storage and the evolving carbon framework \cite{Li2022a}. Collectively, these studies demonstrate that sodium storage is intrinsically controlled by the local carbon environment.

Despite this progress, an important distinction remains between identifying thermodynamically favourable sodium-storage environments and understanding how sodium ions actually access and traverse a heterogeneous carbon electrode. The recent work by Front et al.\cite{Front2026}, for example, demonstrate clear correlations between carbon density, pore accessibility and sodium uptake, but do not explicitly account for percolation pathways, kinetic limitations or electrolyte transport constraints. Similarly, atomistic site-classification approaches provide detailed information on the local environments occupied by sodium but do not directly resolve the three-dimensional trajectories by which solvated sodium ions enter and redistribute within a realistic porous electrode.\cite{Wang2026} These limitations are particularly relevant because maximum penetration into a pore network does not necessarily imply sustained migration or redistribution within the electrode. Establishing this distinction requires explicit treatment of ion trajectories within a structurally heterogeneous three-dimensional carbon network.

Furthermore, the predictive capability of current atomistic simulations remains fundamentally limited by the structural realism of the carbon models employed. Many computational studies, particularly those based on DFT, rely on highly idealised representations such as isolated graphene fragments, slit pores or defect-free graphitic layers that fail to reproduce the interconnected heterogeneity observed experimentally \cite{Li2022,Alptekin2020,Au2020}. Larger-scale atomistic models generated by molecular dynamics provide a significant improvement by naturally producing interconnected porous networks with experimentally relevant pore sizes while allowing structural disorder and local graphitic ordering to be retained. The first realistic atomistic representation of hard carbons, together with their dependence on synthesis temperature, was introduced by de Tomas \textit{et al.}\cite{DeTomas2017} using annealed molecular dynamics and the Environment-Dependent Interaction Potential (EDIP), which accurately reproduces the partial graphitisation characteristic of non-graphitising carbons during high-temperature synthesis \cite{DeTomas2016,DeTomas2017}. More recently, machine-learning interatomic potentials based on the Gaussian Approximation Potential (GAP) framework have enabled the generation of realistic disordered carbon structures with substantially improved accuracy while maintaining computational efficiency at large length scales \cite{Rowe2020,Nanoporous-carbonsTurbogap}. Ongoing developments are progressively extending these models to account for chemical complexity through the incorporation of hydrogen\cite{Ibragimova2025} and oxygen functionalities \cite{Zarrouk2024}. Nevertheless, the use of such structurally realistic machine-learning-generated hard-carbon models to investigate sodium transport in an explicit electrolyte environment remains comparatively limited.

A second and equally important limitation concerns the explicit treatment of the electrolyte. Most atomistic studies investigate sodium insertion either in vacuum or under highly simplified conditions, thereby neglecting the complex interactions occurring at the carbon–electrolyte interface \cite{Lin2025,Front2026,Li2022,Li2022a,Li2022b}. While these approaches have provided valuable insight into the intrinsic storage behaviour of hard carbons, they cannot directly describe electrochemical processes occurring under realistic operating conditions, where sodium ions, solvent molecules and counterions simultaneously interact with the evolving carbon microstructure. Recent work has highlighted the importance of electrolyte access to nanoporous carbon structures, further motivating simulations in which the electrolyte is represented explicitly.\cite{Liu2026} At the same time, no transferable interatomic potential currently provides a fully reactive and computationally tractable description of the complete combination of disordered carbon, sodium, salt and carbonate solvent. Reactive force fields can describe sodium–carbon interactions\cite{Hjertens2016}, while established classical force fields provide efficient descriptions of electrolyte structure and dynamics, but combining these descriptions remains challenging.

To address these challenges, we combine structurally realistic hard-carbon electrodes generated using the GAP-20 machine-learning potential with molecular dynamics simulations of an explicit 1 M NaPF$_6$ electrolyte in an ethylene carbonate/dimethyl carbonate (EC/DMC, 1:1 by volume) solvent mixture. The electrochemical cells are described using a hybrid classical force field incorporating bonded and non-bonded interactions, while constant electrode potentials are imposed using the ELECTRODE implementation in LAMMPS. This framework enables the behaviour of solvated Na$^+$ ions to be examined at the interface with structurally heterogeneous three-dimensional carbon electrodes under applied electrochemical conditions.

Using this approach, we investigate how the coupled evolution of pore accessibility, pore geometry and local graphitic ordering across a systematic density series influences sodium uptake and transport. Rather than assigning storage to independently controlled structural variables or treating adsorption, interlayer insertion and pore filling as mutually exclusive mechanisms, we examine how the resulting carbon microstructure governs sodium penetration, local Na–C and Na–O coordination, and the persistence and geometry of ion trajectories within the electrode. Particular emphasis is placed on distinguishing maximum penetration depth from retained displacement, thereby separating transient exploration of the pore network from sustained inward migration. By explicitly resolving the carbon–electrolyte interface and three-dimensional ion trajectories, the resulting simulations provide an atomistic perspective on how realistic hard-carbon microstructure controls the extent to which electrostatic polarisation translates into sodium penetration and sustained transport.

\section{Methods}

\subsection{Generation of realistic hard carbon electrode models}

Three-dimensional carbon electrode models were generated via a melt-quench-anneal  molecular dynamics protocol using the GAP-20 machine-learned interatomic potential \cite{Rowe2020}  in LAMMPS, implemented in the QUIP interface\cite{Plimpton1995}.  All stages used the canonical (NVT) ensemble,
a Nos\'e--Hoover thermostat with a damping time of 1~ps, and a time step of
2~fs. The simulation box volume did not change during the protocol. Periodic boundary
conditions were applied in all three spatial directions. 
Carbon density was systematically varied from 0.5 and 2.0 g cm$^{-3}$ to generate a series of four nanostructural models exhibiting a progressive evolution in pore accessibility, pore geometry and local graphitic ordering, according to our previous work\cite{DeTomas2018APL}. This was achieved by maintaining the number of carbon atoms constant and varying the simulation box volume (see Table \ref{tab:composition}).
For each density, we scaled one disordered 216-atom configuration to the
target density. We replicated this cubic cell $3\times3\times3$ times to produce
a system with a total of 5,832 atoms.
The box lengths were 61.50, 48.81, 42.64 and 38.74~\AA{} for 0.5, 1.0,
1.5 and 2.0~g\,cm$^{-3}$, respectively (Table~\ref{tab:composition}).
Each simulation had five stages:
(i) Geometry optimisation of the initial structure following volume
scaling. (ii) Atoms were assigned random velocities at 6000~K. The liquid
carbon was kept at 6000~K for 20~ps. (iii) The target temperature was set to 300~K to quench the high temperature liquid. This stage results into an amorphous solid carbon structure. We then increased the target temperature linearly to 3500~K over a 20~ps period. (iv) The amorphous structures were annealed at 3500~K for 200~ps. This high temperature annealing is essential to promote the graphitisation of the amorphous structures~\cite{DeTomas2016}. (v) After annealing, the target temperature was decreased linearly from 3500~K to 300~K in
20~ps. This final stage results into a stable and partially graphitised solid carbon at ambient temperature. 

\subsection{Electrochemical modelling}

\subsubsection*{Construction of atomistic electrochemical cells}

Prior to constructing the electrochemical cells, preparation of the electrodes and the electrolyte was performed separately.  Periodicity was removed from one direction in each hard-carbon electrode creating two free surfaces while the other two dimensions remained periodic. Along the direction normal to the electrode surface. The electrodes were subsequently subjected to an energy minimisation, followed by an equilibration in the NVE ensemble until the surface was reconstructed. 

The electrolyte consisted of a 1 M NaPF$_6$ solution in an ethylene carbonate/dimethyl carbonate (EC/DMC, 1:1 by volume) solvent mixture. Initial electrolyte configurations were generated using Packmol \cite{Martinez2009} and the \textit{fftool} package \cite{Padua2017}, with the salt already dissociated into Na$^+$ ions and PF$_6^-$ anions. 

Electrochemical cells were constructed by sandwiching the electrolyte in between two identical porous hard-carbon electrodes on either side of an electrolyte region, with the free surfaces of the electrodes being in contact with the electrolyte and  forming a symmetric two-electrode cell under periodic boundary conditions.

To maintain a constant salt concentration across carbons with different pore volumes, the number of Na$^+$, PF$_6^-$ and solvent molecules was adjusted according to the electrode density. The corresponding system compositions are summarised in Table~\ref{tab:composition}.

\begin{table}[h]
\centering
\caption{System composition for each electrochemical cell. Model indicates nomenclature of carbon electrode, $d$ refers to the electrode density, and $l$ refers to the electrode face length. Number of electrolyte molecules for each carbon electrode are also indicated.}
\label{tab:composition}
\begin{tabular}{cccccc}
\hline
Model & $d$ & $l$ & \ce{Na+} & \ce{PF6^-} & EC/DMC \\
      & ($\mathrm{g/cm^{3}}$) & (nm) & & & \\
\hline
HC-C1 & 0.5 & 6.15 & 280 & 280 & 2020 \\
HC-C2 & 1.0 & 4.88 & 150 & 150 & 1050 \\
HC-C3 & 1.5 & 4.26 & 100 & 100 & 690  \\
HC-C4 & 2.0 & 3.87 & 80  & 80  & 520  \\
\hline
\end{tabular}
\end{table}

\subsubsection*{Equilibration protocol}

Prior to electrochemical simulations, the assembled electrochemical cells were equilibrated in the isothermal--isobaric (NPT) ensemble at 300 K using a hybrid classical force field to account for interatomic interactions incorporating both bonded and non-bonded terms. Harmonic potentials were employed to describe intramolecular bond stretching and angle bending:
\begin{equation}
E_{\text{bond}} = K_b (r - r_0)^2
\end{equation}
\begin{equation}
E_{\text{angle}} = K_{\theta} (\theta - \theta_0)^2
\end{equation}
where $K_b$ and $K_{\theta}$ denote the force constants, and $r_0$ and $\theta_0$ represent the equilibrium bond lengths and angles, respectively. Non-bonded interactions were represented by a standard 12-6 Lennard-Jones (LJ) potential coupled with a long-range Coulombic term (\texttt{lj/cut/coul/long}):
\begin{equation}
E_{ij} = 4\varepsilon_{ij} \left[ \left(\frac{\sigma_{ij}}{r_{ij}}\right)^{12} - \left(\frac{\sigma_{ij}}{r_{ij}}\right)^6 \right] + \frac{1}{4\pi\varepsilon_0} \frac{q_i q_j}{r_{ij}}
\end{equation}

Cross-interaction LJ parameters ($\sigma_{ij}$ and $\varepsilon_{ij}$) were evaluated using geometric Lorentz--Berthelot mixing rules (\texttt{pair\_modify mix geometric}), and analytical tail corrections were applied to account for long-range dispersive energy contributions (\texttt{tail yes}). Short-range pairwise interactions were truncated at a cutoff radius of \SI{12.0}{\angstrom}. Long-range electrostatic interactions were calculated using the Particle-Particle Particle-Mesh (PPPM) solver with a relative force accuracy threshold of $1.0 \times 10^{-4}$. Intramolecular 1--2 and 1--3 non-bonded interactions were fully scaled to zero, whereas 1--4 interactions were scaled by a factor of 0.5 (\texttt{special\_bonds lj/coul 0.0 0.0 0.5}). Non-bonded interaction parameters assigned to each atom type are summarized in the Supplementary Information.

Periodic boundary conditions were applied in all three spatial dimensions ($x, y, z$). Prior to production dynamics, energy minimization was performed using the Conjugate Gradient algorithm down to an energy stopping tolerance of $1.0 \times 10^{-4}$ (unitless) and a force stopping tolerance of $1.0 \times 10^{-6}$~\si{kcal.mol^{-1}.\angstrom^{-1}}.

Atomic velocities were randomly assigned from a Maxwell--Boltzmann distribution corresponding to $T = \SI{300.0}{K}$. Equations of motion were integrated using the velocity Verlet algorithm with a timestep of $\Delta t = \SI{1.0}{fs}$. Neighbor lists were constructed using a binning scheme with a skin distance of \SI{2.0}{\angstrom} (\texttt{neighbor 2.0 bin}).

The NPT equilibration was conducted using a Nosé--Hoover thermostat and barostat. System temperature and pressure were maintained at $T = \SI{300.0}{K}$ and $P = \SI{1.0}{bar}$ with damping parameters of $T_{\text{damp}} = \SI{100}{fs}$ ($100\Delta t$) and $P_{\text{damp}} = \SI{500}{fs}$ ($500\Delta t$), respectively. Isotropic pressure coupling was applied across all spatial directions. 
  Pressure was controlled only along the direction normal to the electrode surfaces ($z$ axis), allowing the electrode separation to relax while maintaining the lateral cell dimensions fixed. During this stage, the carbon electrodes were treated as rigid and electrically neutral, whereas all electrolyte species evolved freely.

A timestep of 0.1 fs was employed during pressure equilibration. Simulations were continued until the pressure along the $z$ direction reached a stable value, which happened at around 10 ns.  During equilibration, the electrolyte wetted the electrodes and the final equilibrium sodium population inside the electrodes was therefore recorded for each electrode and subsequently used as the reference state $N^0_Na$ when determining the net sodium insertion during charging. 

\subsubsection*{Constant-potential molecular dynamics simulations}

Electrochemical charging simulations were subsequently performed using the constant-potential implementation of the ELECTRODE package coupled to LAMMPS \cite{AhrensIwers2022,Plimpton1995}. Equal and opposite electrical potentials were imposed on the two electrodes using the \texttt{fix electrode/conp} algorithm under symmetric boundary conditions. Electrode atomic charges were represented by Gaussian charge distributions whose magnitudes were determined self-consistently at every timestep to satisfy the imposed constant-potential constraint.

Throughout the production simulations, the carbon electrodes remained rigid, whereas the electrolyte evolved under the canonical (NVT) ensemble at 300 K using a Nosé--Hoover thermostat applied exclusively to the electrolyte.

Bonded interactions were described using harmonic bond and angle potentials together with OPLS dihedral potentials. Non-bonded interactions were represented using the hybrid \texttt{lj/cut/coul/long} potential with a cutoff distance of 12~Å for both Lennard--Jones and real-space Coulombic interactions. Lennard--Jones interactions between unlike atom types were generated using geometric mixing rules, whereas long-range electrostatic interactions were computed using the PPPM/electrode solver with a relative accuracy of $10^{-4}$. Bond constraints were enforced using the SHAKE algorithm. All parametrizations are detailed in the Supplementary Information.

Production simulations employed a timestep of 1 fs and were propagated for 20 ns, with atomic configurations written every 50 ps. Electrochemical charging simulations were performed at applied electrode potentials of 0.1, 0.2, 0.4, 0.5, 1.0, 1.5, 2.0, 2.5 and 3.0 V and the incremental uptake of sodium ions inside the electrodes was recorded as $\Delta N_{\rm Na}(V)=N_{\rm Na}(V)-N_{\rm Na}^{0}$. 

\subsection{Structural and electrochemical characterisation}

\begin{figure}[tb]
    \centering
    \includegraphics[width=\columnwidth]{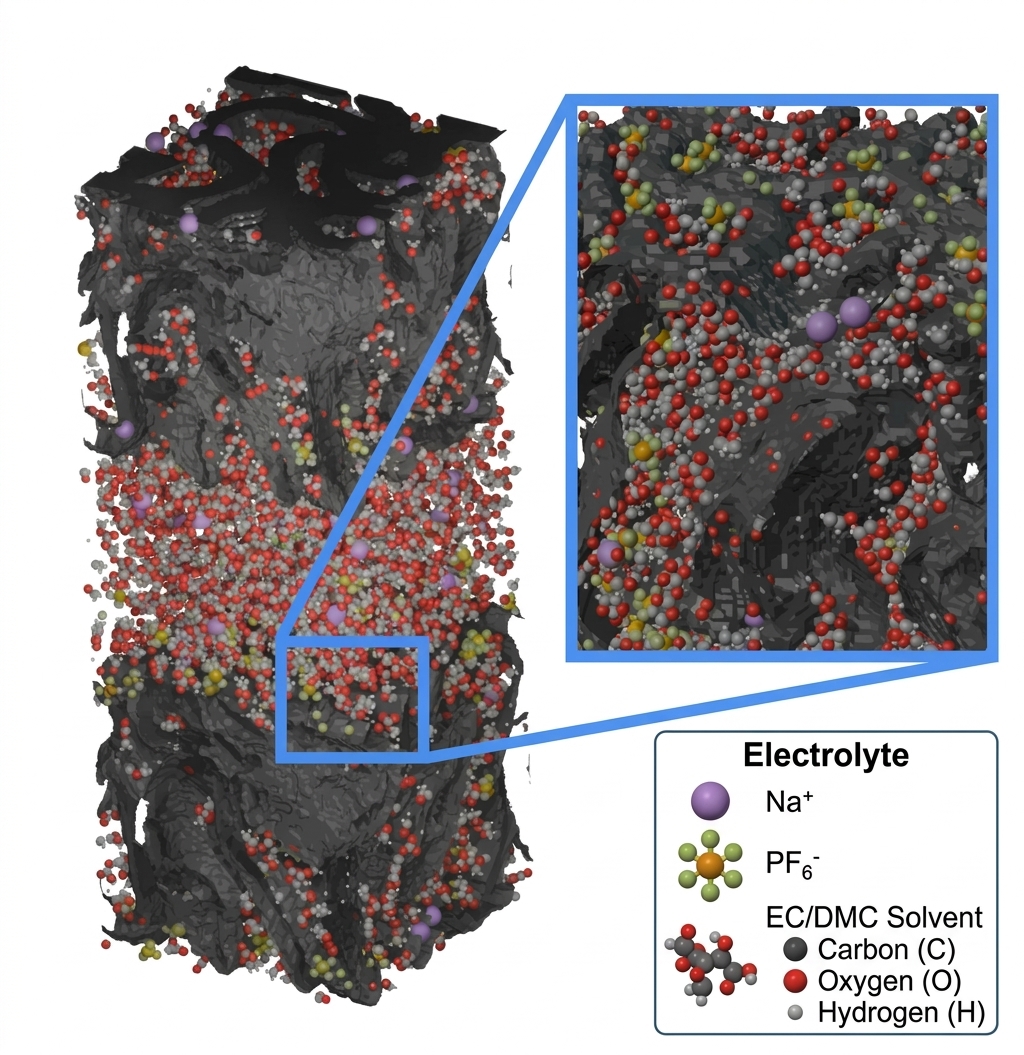}
    \caption{Snapshot of the atomistic electrochemical simulation cell comprising two hard carbon electrodes (black nanostructures) separated by a \ce{NaPF6} in EC:DMC electrolyte. The snapshot corresponds to the charge equilibration final stage of the HC-C2 electrodes at 0.1 V of applied voltage. The surface of the electrodes nanostructure are rendered to aid visualisation of the porous network, and the electrolyte molecules are coloured according to the legend. A zoom-in view of the highlighted electrode area in the blue frame is shown on the right panel.}
    \label{fig:Simulation_Box}
\end{figure}

\subsubsection*{Carbon electrode microstructure characterisation}

The microstructure of the generated hard-carbon electrode models was characterised by simulated X-ray diffraction (XRD) patterns, high-resolution transmission electron microscopy (HR-TEM) images, porosity analysis and calculation of the specific surface area (SSA).

X-ray diffraction patterns were calculated using the Debye scattering equation~\cite{Warren1969}

\begin{equation}
I(s)=
\sum_{i=1}^{N}
\sum_{j=1}^{N}
f_i(s)f_j(s)
\frac{\sin(2\pi s r_{ij})}
{2\pi s r_{ij}},
\end{equation}

where $s$ is the magnitude of the scattering vector, $r_{ij}$ is the distance between atoms $i$ and $j$, $N$ is the total number of atoms, and $f(s)$ is the atomic scattering factor of carbon.

High-resolution transmission electron microscopy images were simulated using the multislice method of Kirkland~\cite{kirkland2016}, implemented in the \textit{compuTEM} software, employing an accelerating voltage of 80 kV and a focal depth corresponding to the thickness of the simulation box. Porosity characterisation was performed with Poreblazer~\cite{Sarkisov2020} using a nitrogen probe to obtain N$_2$ accessible specific surface area (SSA) and accessible porosity, whereas a helium probe was employed to geometrically quantify the maximum pore size in each model.
 
 Local atomic structure was characterized through coordination numbers, ring topology, and Smooth Overlap of Atomic Positions (SOAP) descriptors. Coordination numbers were calculated using a nearest-neighbor cutoff of 1.85~\AA. Ring statistics were determined via Franzblau's shortest-path algorithm~\cite{Franzblau}. Smooth Overlap of Atomic Positions (SOAP)
descriptors\cite{Bartok2013} were used to
compare the local atomic environments in the models with three reference
structures: AB-stacked graphite, the C$_{60}$ fullerene and a
($9$,$9$) single-wall carbon nanotube. SOAP parameters were selected to match those used to fit the GAP-20 potential itself. We used a cutoff radius
$r_\mathrm{c}$ = 6.0~\AA{}, a cutoff transition
width of 1.0~\AA{}, a Gaussian width $\sigma_\mathrm{at}$ = 0.5~\AA{},
$n_\mathrm{max}$ = 12 and $l_\mathrm{max}$ = 4. Reference structures were relaxed using the GAP-20 potential prior to performing the calculation.
For each atom $i$, we calculated the SOAP vector $\mathbf{p}_i$ and normalised it to unit length, $\hat{\mathbf{p}}_i$. For each structure $A$ with $N_A$
atoms, we calculated the average vector
$\bar{\mathbf{p}}_A = N_A^{-1}\sum_{i\in A}\hat{\mathbf{p}}_i$. The similarity $K$
between a model $A$ and a reference structure $B$ is the normalised dot product of
their average vectors:
\begin{equation}
K(A,B) = \left(
\frac{\bar{\mathbf{p}}_A \cdot \bar{\mathbf{p}}_B}
{\lvert\bar{\mathbf{p}}_A\rvert \,
\lvert\bar{\mathbf{p}}_B\rvert}
\right)^{\zeta},
\end{equation}
with $\zeta$ = [1]. $K = 1$ when two structures have the same distribution
of local
environments. $K$ decreases as the environments become different. 
The cutoff radius is larger than the interlayer distance of graphite (3.35~\AA{}). Thus, the similarity to graphite includes both the effect of curvature and the effect of interlayer stacking.
   
\subsubsection*{Electrolyte characterisation}

Partial radial distribution functions (RDFs) and coordination numbers (CNs) were calculated from the production LAMMPS trajectories using custom Python analysis scripts. Pairwise distances were evaluated under periodic boundary conditions using the minimum-image convention and accumulated into radial histograms over all simulation frames. RDFs were obtained by normalising the pair distributions by the corresponding spherical shell volumes and the average number density of the second species.

Partial RDFs were calculated for Na--O atom pairs in order to characterise sodium solvation by solvent molecules, and Na--C atom pairs to characterise the interaction between sodium and the carbon electrode, respectively. Coordination numbers were determined by integrating the corresponding RDFs up to the first minimum following the first coordination shell, which was identified automatically from each distribution.

\subsubsection*{Electrode pore connectivity and electrolyte accessibility}

To analyse pore connectivity within the electrodes and electrolyte accessibility, the simulation cell was discretised into a regular three-dimensional voxel grid with a spacing of 0.6~\AA. Carbon atoms belonging to the upper and lower electrodes were identified from their molecule IDs and treated independently to construct binary solid masks for each electrode. Voxels whose centres were located within an exclusion radius of 2.2~\AA\ from any carbon atom were classified as belonging to the solid phase. Before voxelisation, carbon atoms were re-imaged into the corresponding half of the simulation cell to avoid artificial overlap between the two electrodes arising from periodic boundary conditions. The individual electrode masks were subsequently combined to define the complete solid phase. The remaining void space was classified into exterior electrolyte and internal pore regions using a flood-fill algorithm applied to the voxel grid. Void voxels connected to the boundaries of the simulation cell were identified as exterior electrolyte, whereas disconnected void regions were classified as internal pores. The resulting solid, pore and exterior masks were exported for subsequent visualisation and transport analyses.

Continuous \ce{Na+} trajectories were reconstructed from the molecular dynamics simulations by unwrapping periodic boundary crossings using the minimum-image convention. Frame-to-frame atomic displacements were accumulated to generate continuous trajectories throughout the complete production simulations. The reconstructed trajectories were combined with the voxelized pore masks to identify preferential Na-ion entry locations and migration pathways within the porous carbon electrodes. Representative trajectories were tracked and visualised in MATLAB together with the reconstructed pore network to investigate pore accessibility and infiltration behaviour.

The reconstructed trajectories were used for displacement calculations, whereas coordinates mapped to the primary simulation cell were used to determine the position of the ions relative to the electrode–electrolyte interfaces.

The retained net displacement of \ce{Na+} along the electrode normal was quantified from the absolute net z-displacement of each ion,

\[ 
\Delta z_i = |z_i(t_f)-z_i(t_0)|
\] 

where $z_i(t_0)$ and $z_i(t_f)$ are the initial and final unwrapped $z$-coordinates of \textit{i-th} \ce{Na+}. This quantity measures the component of the ion displacement retained at the end of the analysed trajectory and therefore distinguishes sustained redistribution along the electrode normal from local motion followed by a return towards the initial position. The sodium ions exhibiting the largest values of $|\Delta z_i|$ were additionally visualised within the three-dimensional carbon structure to identify representative long-range transport pathways. 

Because net displacement alone does not capture transient excursions into the porous electrodes, the maximum penetration depth was determined independently for every \ce{Na+}. The electrolyte-facing surfaces of the upper and lower electrodes were identified from the voxelised carbon structures. For each trajectory frame, the distance travelled by the \ce{Na+} beyond the corresponding electrolyte–electrode interface and towards the interior of either electrode was calculated. The maximum value sampled during the trajectory was defined as the maximum penetration depth,

\[ 
P_i^{max} = \underset{t}{max},\ P_i(t),
\]

where $P_i(t)$ is the instantaneous penetration depth of ion \textit{i}. Consequently, $P_i^{max} =0$ denotes an ion that did not enter either electrode according to this geometrical definition, whereas increasing $P_i^{max}$ indicates progressively deeper exploration of the porous carbon. This definition is independent of the ion's final position and therefore captures penetration events even when an ion subsequently returns towards the electrolyte.

The geometrical character of each trajectory was further quantified using the trajectory straightness,

\[ 
S_i = \frac{|\mathbf{r}_i(t_f) - \mathbf{r}_i(t_0)|}{\sum^{N-1}_{j=1} |\mathbf{r}_i(t_{j+1}) - \mathbf{r}_i(t_j)|}
\]

where the numerator is the three-dimensional net displacement and the denominator is the cumulative three-dimensional path length. $S_i$ approaches unity for highly directed motion and decreases towards zero for increasingly tortuous trajectories. Straightness was used as a complementary descriptor rather than as a direct measure of mobility, since a short trajectory can be highly straight without corresponding to substantial ion transport.

    \begin{figure*}[tb]
    \centering
    \includegraphics[width=0.75\linewidth]{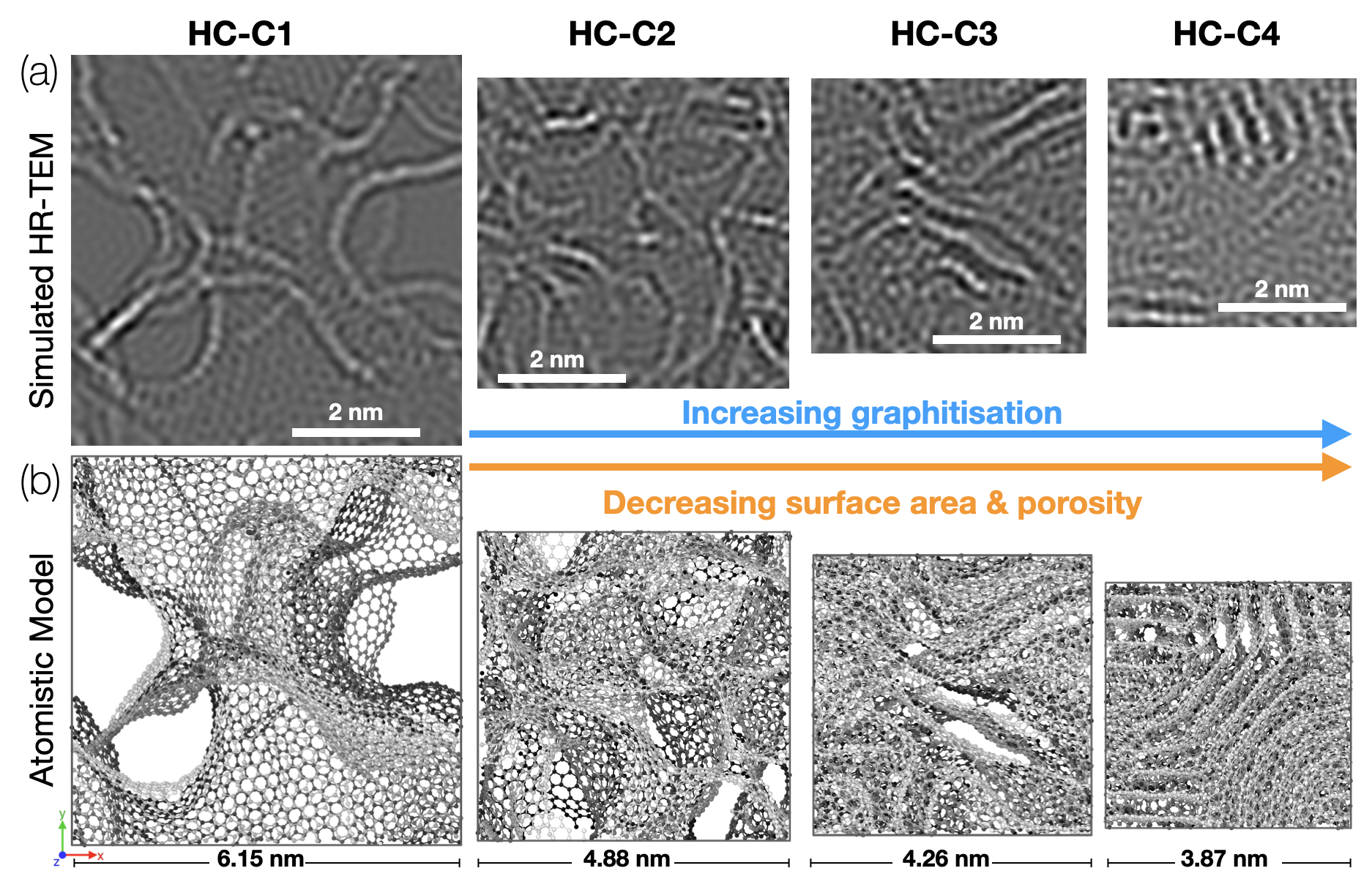}
    \caption{Atomistic carbon models HC-C1 through HC-C4 with increasing densities of 0.5, 1.0, 1.5, and 2.0 g cm$^{-3}$, respectively. Top panels (a) show simulated HR-TEM images of the atomistic nanostructures rendered below (bottom panels (b)). Models depict the x–y plane exposed to the electrolyte. Simulation box dimensions are labeled below each model, and all images share a 2 nm scale bar. }
    \label{fig:HC_models}
\end{figure*}

To relate the trajectory-based analysis to the thermodynamic landscape sampled by \ce{Na+}, one-dimensional Na-ion probability density profiles, $\rho(z)$, were calculated along the electrode normal using all production frames. The z-coordinate was discretised into uniform bins and the accumulated Na-ion counts were normalised by the bin volume and number of analysed frames. The resulting probability density was converted into a potential of mean force (PMF) according to

\[
\Delta F(z)
=
-k_{\mathrm{B}}T
\ln\left(
\frac{\rho(z)}
{\rho_{\mathrm{ref}}}
\right),
\]

where $\rho(z)$ is the Na-ion probability density at position $z$, $\rho_{\mathrm{ref}}$ is a reference density, $k_{\mathrm{B}}$ is Boltzmann's constant and $T$ is the simulation temperature. Because only relative free energies are relevant, the resulting PMF was shifted such that its minimum corresponded to zero. Regions of low $F(z)$ therefore correspond to positions sampled with comparatively high probability, whereas larger values indicate less frequently sampled regions along the projected transport coordinate.

For each \ce{Na+} exhibiting pore penetration, the PMF was evaluated at both its initial position, $z_i(t_0)$, and the position at which its maximum penetration occurred, $z_i^{\mathrm{deep}}$. The free-energy change associated with reaching the deepest sampled position was consequently defined as

\[
\frac{\Delta F_i^{\mathrm{deep}}}{k_{\mathrm{B}}T} = \ln \left[ \frac{\rho[z_i(t_0)]}{\rho(z_i^{\mathrm{deep}})}  \right]
\]

Negative values of $\Delta F_i^{\mathrm{deep}}$ indicate that the position corresponding to maximum penetration is more highly populated in the projected PMF than the initial position, whereas positive values indicate penetration into a comparatively less populated region. $\Delta F_i^{\mathrm{deep}}$ represents a free-energy difference derived from the ensemble-averaged one-dimensional probability distribution and should not be interpreted as the microscopic binding energy of a specific pore or adsorption site.





\begin{figure}[tb]
    \centering
    \includegraphics[width=0.9\columnwidth]{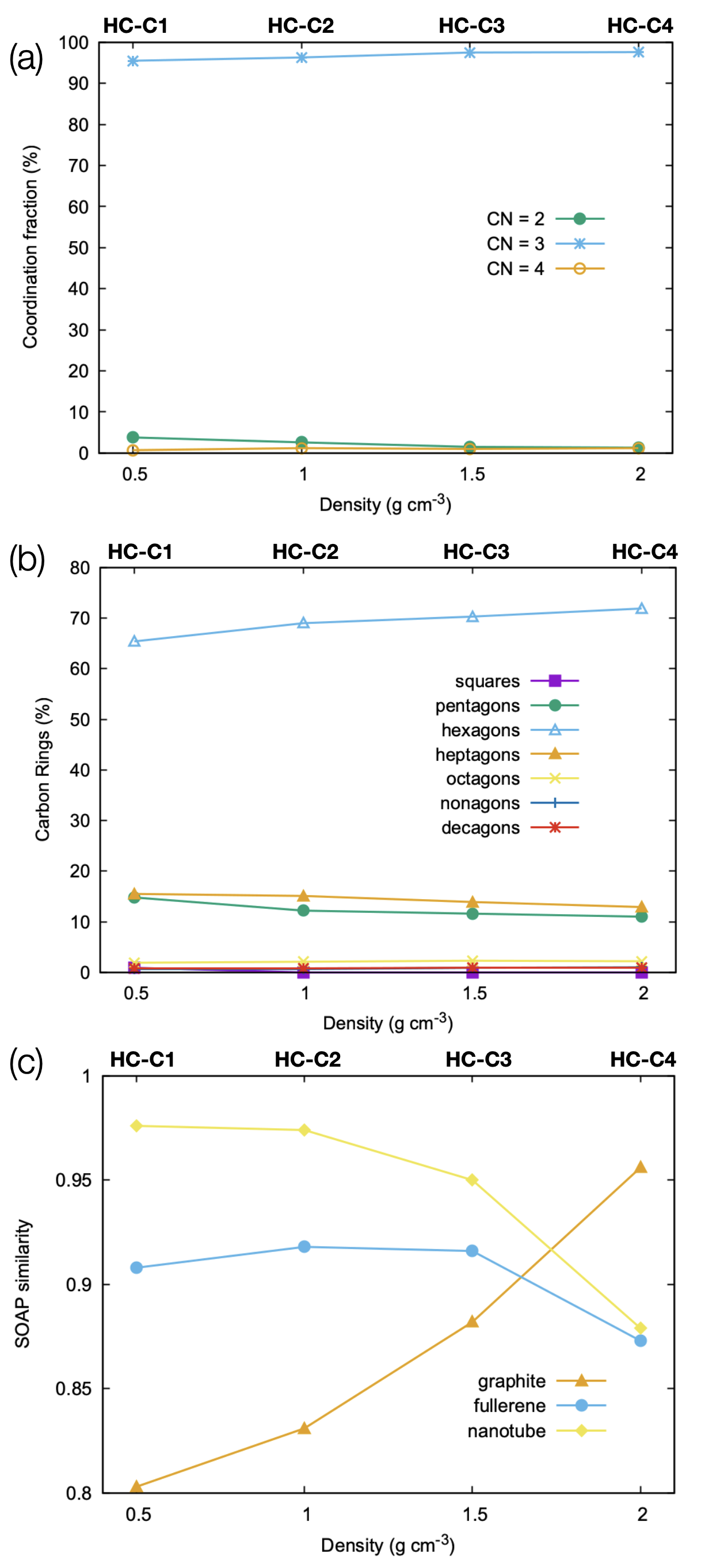}
    \caption{Nanostructure characterisation of the carbon models  HC-C1 through HC-C4 with densities of 0.5, 1.0, 1.5 and 2.0 g\,cm$^{-3}$ respectively. a) Relative fractions of 2-, 3-, and 4-coordinated carbon atoms, b) ring-size distribution from 4- to 10-membered rings (squares to decagons), and c) Smooth Overlap of Atomic Positions (SOAP) structural similarity to ideal graphite, fullerenes and nanotubes.}
    \label{fig:HC_characterisation}
\end{figure}

\section{Results and discussion}

\subsection{Microstructure Characterisation of Carbon Electrode Models}

The generated carbon electrode models, HC-C1 to HC-C4, span a broad range of porosities and structural ordering, while retaining the topological disorder characteristic of hard carbon. The free surfaces of the four models are shown in Fig.~\ref{fig:HC_models}(b), and the corresponding local structural descriptors are summarised in Fig.~\ref{fig:HC_characterisation}. All models are predominantly $sp^2$-hybridised (\textit{i.e.}, 3-coordinated carbon atoms), with the fraction of $sp^2$ carbon  increasing only slightly from 95.5\% at 0.5~g~cm$^{-3}$ (HC-C1)to 97.6\% at 2.0~g~cm$^{-3}$ (HC-C4).  The remaining carbon is mainly $sp$- or $sp^3$-hybridised, with the $sp$ fraction decreasing from 3.8\% to 1.3\% over the same density range and $sp^3$ fluctuating around 1\% for all models. Thus, the systematic evolution of the microstructure with density does not arise from a substantial change in the overall hybridisation state, but rather from changes in the topology and local ordering of the predominantly $sp^2$ carbon network. 

The ring statistics reveal a predominance of six-membered rings (hexagons) in all models, which account for 65.4--71.9\% of the identified rings. Pentagons and heptagons constitute most of the remaining population, with their combined fraction decreasing from 30.3\% at 0.5~g~cm$^{-3}$ to 24.0\% at 2.0~g~cm$^{-3}$. Smaller populations of eight-, nine- and ten-membered rings are present in all models, while four-membered rings are absent except at the lowest density (0.9\%) and three-membered rings are not present. The reduction in the population of non-hexagonal rings with increasing density is consistent with the progressive graphitisation of the carbon network. These topological defects nevertheless remain present even in the densest model, indicating that the structures retain a significant degree of disorder rather than evolving towards ideal crystalline graphite.

A clear density-dependent evolution is also observed in the morphology of the carbon network. At low density, the structures form highly porous and percolated networks with strongly curved pore walls, with pore geometries being primarily amorphous voids. As the density increases, the pore network progressively contracts and the carbon walls become flatter, accompanied by the emergence of increasingly ordered graphitic domains which results in a predominance of slit-like pore geometries. The development of graphitic character is particularly evident in the simulated HR-TEM images (Fig.~\ref{fig:HC_models}(a)). The lowest-density structure, HC-C1, contains highly curved carbon walls with no discernible graphitic stacking. Its pore network is predominantly microporous, although a small mesoporous contribution is present, with a maximum pore diameter of 3.04~nm (Table~\ref{tab:HC_samples}). With increasing density, the pore walls become progressively flatter and the pore dimensions decrease. At 1.5~g~cm$^{-3}$ (HC-C3), the first graphitic ribbons become apparent, giving rise to slit-like pores. At the highest density, HC-C4 contains larger, defective and randomly oriented graphitic stacks. Thus, the transition from HC-C1 to HC-C4 represents a progressive evolution from a highly curved and porous carbon network towards a denser structure containing extended, although still disordered, pseudographitic domains.


The visual evolution observed in the HR-TEM images is quantitatively supported by the SOAP descriptors. The SOAP similarity to ideal graphite increases systematically with density, from 0.803 for HC-C1 to 0.956 for HC-C4, despite only a modest increase in the overall $sp^2$ fraction (95.5–97.6\%). This demonstrates that the density-driven increase in graphitic character arises primarily from a reorganisation of the predominantly $sp^2$ carbon network towards more graphite-like local environments, rather than from a substantial change in carbon hybridisation. The increase in graphite similarity becomes particularly pronounced between 1.5 and 2.0 g cm$^{-3}$, coinciding with the first appearance of graphitic ribbons in HC-C3. This structural transition is consistent with our previous simulations of non-graphitising carbons using classical interatomic potentials, which predicted the onset of graphitic stacking at a similar density of 1.5 g cm$^{-3}$.\cite{DeTomas2017,DeTomas2016,DeTomas2018APL} Here, the SOAP analysis provides a quantitative measure of this transition and demonstrates that the same density-dependent structural evolution is reproduced using the GAP-20 machine-learning interatomic potential. In contrast, the SOAP similarity to nanotube environments decreases from 0.976 to 0.879 with increasing density, while the fullerene similarity decreases from 0.908–0.918 at lower densities to 0.873 at 2.0 g cm$^{-3}$, consistent with the progressive flattening of the carbon walls. Together, these trends show that SOAP can distinguish the transition from highly curved $sp^2$-carbon environments at low density to increasingly planar, graphite-like environments at high density, providing a quantitative descriptor of graphitic ordering in the generated carbon networks.
\begin{table*}[t]

\caption{Probe-accessible specific surface area (SSA), porosity (P) and maximum pore size ($d_{\rm max}$) calculated using a N$_2$ probe for the four carbon electrodes models.}
\label{tab:HC_samples}
\begin{tabular}{c c c c l}
\hline
 Model & SSA  & P  & $d_{\rm max}$  &Pore geometry \\
  & (m$^2$/g) & (\%) & (nm) & \\
\hline
  HC-C1 & 2351 & 76.7 & 3.04 & amorphous micro-mesopores \\
  HC-C2 & 868  & 37.8 & 1.63 & amorphous micropores \\
  HC-C3 & 0    & 0.0 & 0.94 &  slit-like micropores \\
  HC-C4 & 0    & 0.0 & 0.49 & slit-like ultranarrow micropores \\
\hline
\end{tabular}
\end{table*}

The pore structure evolves concurrently with this increase in structural ordering. The specific surface area and the maximum pore size decrease with increasing density, reflecting the progressive reduction and narrowing of the accessible pore network. In particular, the HC-C3 and HC-C4 structures show no accessible surface area to $N_2$, while geometrically their maximum pore diameters decrease to 0.94 and 0.49~nm, respectively (Table~\ref{tab:HC_samples}). The resulting evolution therefore encompasses a transition from a highly porous network containing both micro- and small mesopores at low density to a predominantly ultramicroporous structure at higher density, accompanied by the development of graphitic domains.

\begin{figure}[t]
    \centering
    \includegraphics[width=\columnwidth]{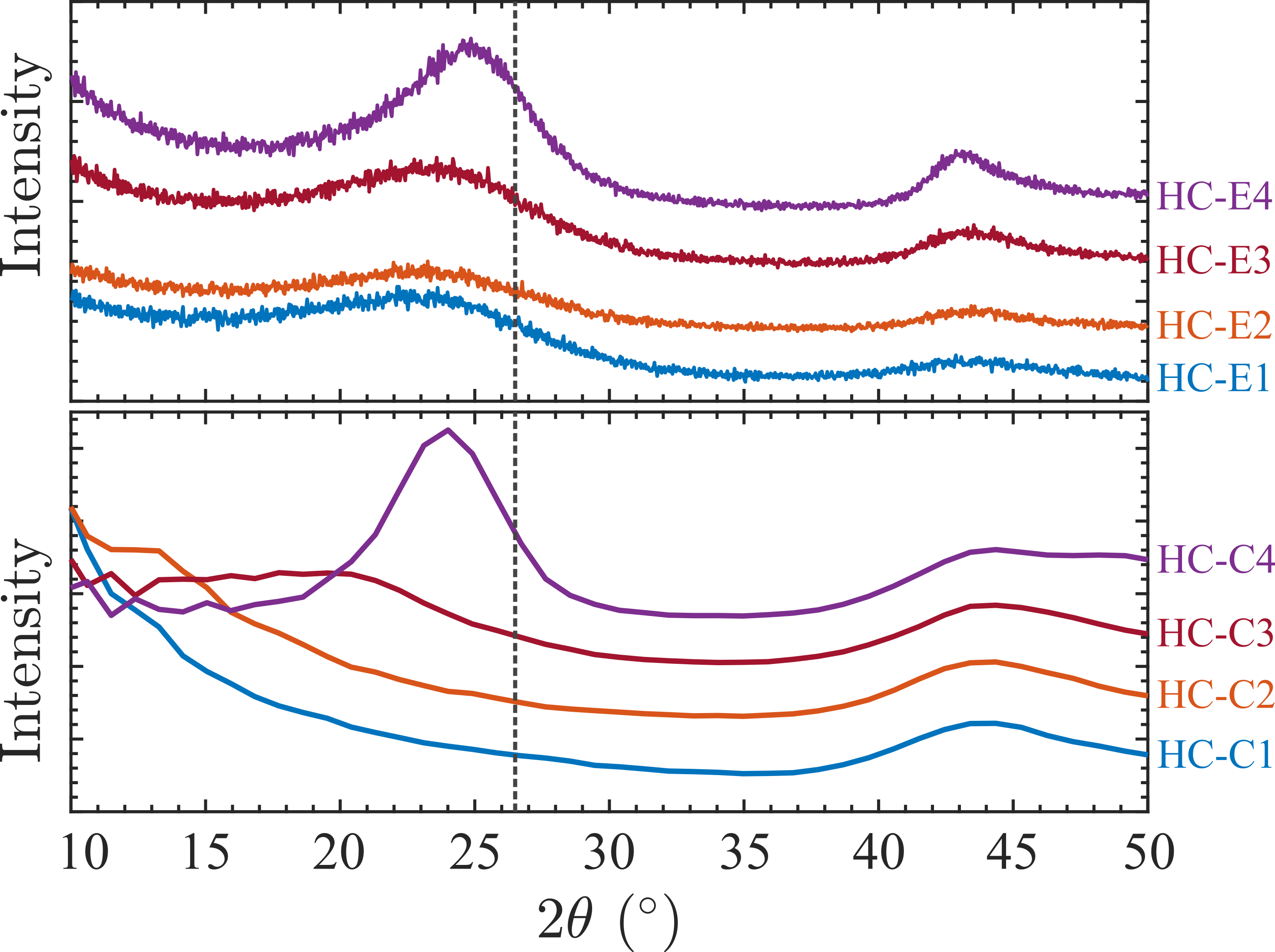}
    \caption{Experimental and simulated diffraction profiles. (a) Experimental XRD patterns for a series of hard-carbon samples with increasing carbonisation conditions (HC-E1 to HC-E4), data replotted from \cite{Guo-Titirici2023}. (b) Simulated XRD patterns for the carbon electrode models from HC-C1 and HC-C4. The dashed line at $2\theta = 26.5^\circ$ marks the position of the 002 peak reflection in ideal graphite.}
    \label{fig:xrd}
\end{figure}

The emergence of stacking is further supported by the simulated XRD patterns (Fig.~\ref{fig:xrd}). The denser models exhibit a broad $(002)$ reflection, which occurs at lower $2\theta$ than the corresponding reflection of ideal graphite. The lower diffraction angle indicates an increased interlayer spacing within the graphitic domains, consistent with the turbostratic and defective nature of the simulated structures. The $(002)$ feature progressively weakens with decreasing density and disappears for the lowest-density structures, in agreement with the absence of discernible stacking in the HR-TEM images. The XRD results therefore provide an independent structural signature of the density-driven transition towards pseudographitic ordering.

Overall, the characterisation results demonstrate that increasing density systematically drives the carbon networks from highly porous, strongly curved and topologically disordered structures towards denser networks containing increasingly graphite-like local environments and pseudographitic stacking. Importantly, this evolution occurs while the models remain predominantly $sp^2$-hybridised and retain non-hexagonal rings and other topological defects characteristic of hard carbon. The observed trends are qualitatively consistent with experimental hard carbons\cite{Guo-Titirici2023} (see Fig.\ref{fig:xrd}a), where increasing carbonisation conditions promotes the development and growth of pseudo-graphitic domains, accompanied by increased stacking and reduced interlayer spacing.\cite{DeTomas2017,DeTomas2016,DeTomas2018APL} Thus, the four models provide a controlled structural series in which increasing density simultaneously reduces pore sizes and accessible surface area, and promotes  local graphitic ordering.  This provides a suitable framework for systematic correlation with the penetration and coordination of \ce{Na+} under an applied potential.

\subsection{Potential-dependent \ce{Na+} uptake and electrode charging}

\begin{figure}[tb]
    \centering
    \includegraphics[width=0.9\columnwidth]{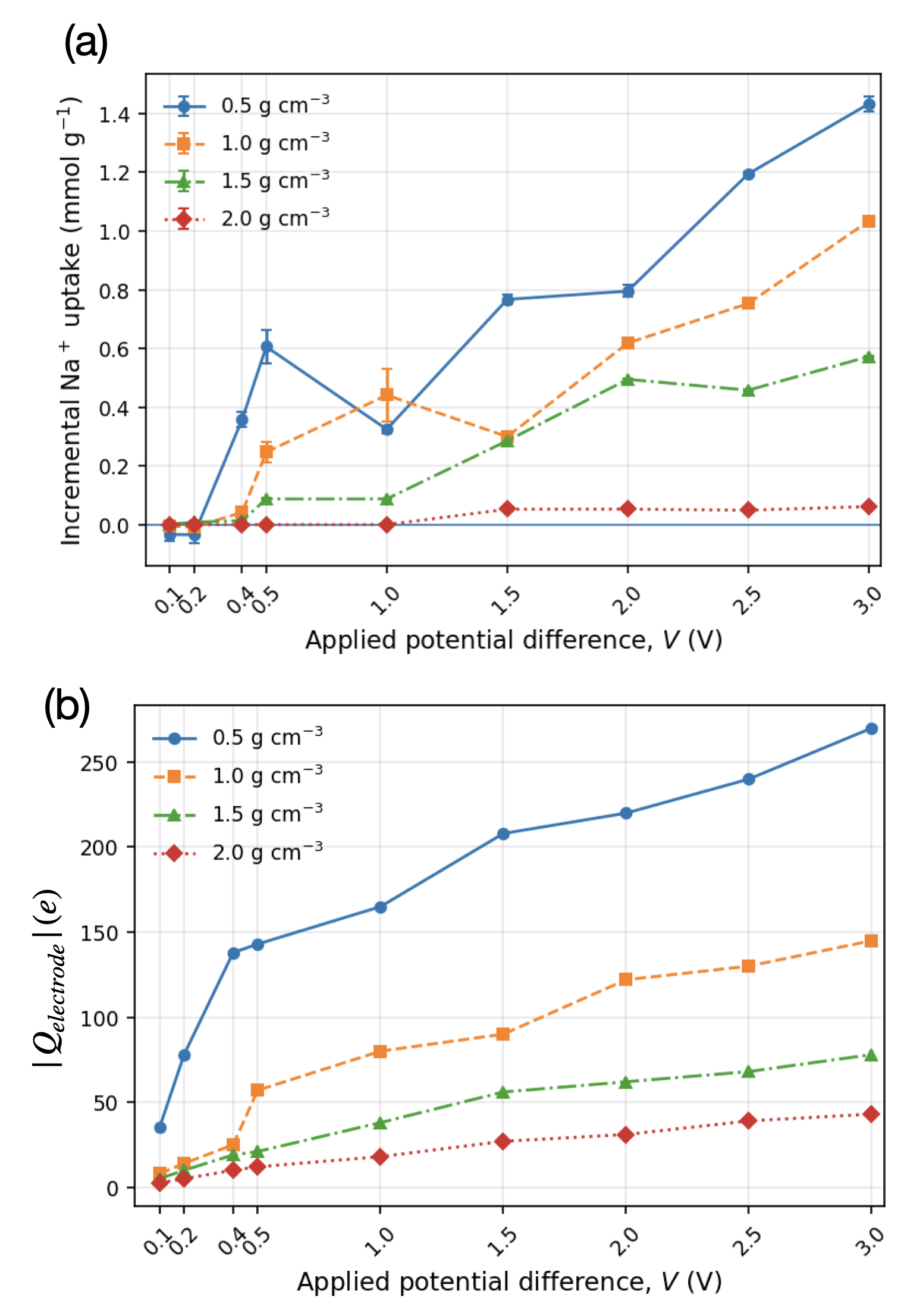}
    \caption{Potential-dependent sodium uptake and electrode charging in the carbon electrode models HC-C1–HC-C4. (a) Incremental \ce{Na+} uptake relative to the initial number of \ce{Na+} ions present in each carbon electrode, normalised by the carbon electrode mass. Error bars represent the uncertainty associated with the averaged \ce{Na+} counts. (b) Magnitude of the electrode charge, $|Q_{\mathrm{electrode}}|$, versus applied potential in each carbon electrode.}
    \label{fig:charge}
\end{figure}

To examine how the carbon microstructure influences the electrochemical response, the number of \ce{Na+} ions located inside the carbon electrodes was analysed as a function of the applied potential. The computational systems were subjected to potential differences ranging from 0.1 to 3.0 V, and the \ce{Na+} population was obtained from the equilibrated portions of the trajectories. The simulations also provided the total charge accumulated on the carbon electrodes, allowing the relationship between electrostatic charging and \ce{Na+} penetration to be assessed independently. The charge-versus-time trajectories were fitted using an exponential relaxation function, as shown in the Supplementary Information. Because the four carbon models contained different numbers of \ce{Na+} ions at the beginning of the simulations, the total \ce{Na+} population cannot be used directly to quantify voltage-induced uptake. Instead, the change in the number of \ce{Na+} ions relative to the initial population was calculated as $\Delta N_{\mathrm{Na}}=N_{\mathrm{Na}}(V)-N_{\mathrm{Na}}(V_{\mathrm{initial}})$ and normalised by the carbon electrode mass. The resulting quantity is reported in Fig.~\ref{fig:charge} incremental sodium uptake in $\mathrm{mmol\,g^{-1}}$. This definition separates the sodium population inherited from the initial configuration from the additional uptake associated with the applied potential. 
As shown in Fig.~\ref{fig:charge}(a), sodium uptake depends strongly on both the applied potential and the carbon density. The lowest-density model, HC-C1, exhibits the largest increase in \ce{Na+} population over the investigated potential range. Its incremental uptake rises from approximately zero at 0.1–0.2 V to around 1.43\ $\mathrm{mmol\,g^{-1}}$ at 3.0 V. HC-C2 displays a similar potential-dependent increase, although the uptake remains lower than that of HC-C1, reaching approximately 1.03\ $\mathrm{mmol\,g^{-1}}$ at 3.0 V. The denser models show substantially lower uptake. HC-C3 reaches approximately 0.57\ $\mathrm{mmol\,g^{-1}}$, whereas HC-C4 exhibits only a small increase, reaching approximately 0.06\ $\mathrm{mmol\,g^{-1}}$ at the highest applied potential.

The density dependence is particularly pronounced at low and intermediate potentials. For HC-C4, essentially no additional \ce{Na+} enters the carbon structure below approximately 1.5 V. In contrast, HC-C1 and HC-C2 begin to exhibit measurable uptake at lower potentials, consistent with the greater pore accessibility and surface area of the less dense structures. These results indicate that the applied electrostatic driving force does not translate directly into \ce{Na+} penetration. Instead, the extent of uptake is additionally controlled by the accessibility and connectivity of the internal carbon network. The uptake curves are not strictly monotonic at every potential. For example, HC-C1 shows a decrease between 0.5 and 1.0 V, HC-C2 decreases between 1.0 and 1.5 V, and HC-C3 decreases slightly between 2.0 and 2.5 V. These deviations are small relative to the overall density-dependent trend and may reflect finite sampling, or fluctuations in the number of ions crossing the electrode boundary. The 0.5 V simulations are particularly sensitive to equilibration, as the corresponding \ce{Na+} populations exhibit comparatively slow and oscillatory evolution in the trajectories. Consequently, individual points should not be interpreted as independent thermodynamic equilibrium capacities.

The electrode charging behaviour is shown in Fig.~\ref{fig:charge}(b). The magnitude of the accumulated electrode charge increases with applied potential for all four carbon densities, although the magnitude of the response decreases systematically as the carbon density increases. HC-C1 exhibits the largest electrode charge throughout the potential range, followed by HC-C2, HC-C3 and HC-C4. This ordering is consistent with the greater ability of the more open carbon structures to accommodate charge redistribution at the carbon–electrolyte interface. The charge curves also show that the electrodes respond electrostatically even under conditions where little or no additional \ce{Na+} penetration is observed.

This distinction is especially clear for HC-C4. Although its electrode charge increases with applied potential, the corresponding \ce{Na+} uptake remains close to zero up to approximately 1.0 V and remains very small at higher potentials. Thus, electrode polarisation alone is insufficient to produce substantial \ce{Na+} storage in the densest carbon model. The applied potential must be accompanied by accessible transport pathways through which \ce{Na+} can enter and populate the carbon network.  Electrode charging and \ce{Na+} uptake are coupled but non-equivalent processes. The charge response reflects the electrostatic polarisation of the electrode, whereas the \ce{Na+} uptake additionally depends on ion transport, pore accessibility and confinement within the carbon structure. Lower-density carbons support both stronger charging and greater \ce{Na+} penetration, while the denser models can sustain an electrostatic response without significant ionic uptake. This provides a direct link between the electrodes structural trends and the microscopic storage mechanisms discussed below.

\subsection{Mechanisms of Na storage}

\subsubsection*{Evolution of Na–C (electrode) and Na-O (electrolyte) coordination numbers with applied potential}

\begin{figure}[tb]
    \centering
    \includegraphics[width=0.8\columnwidth]{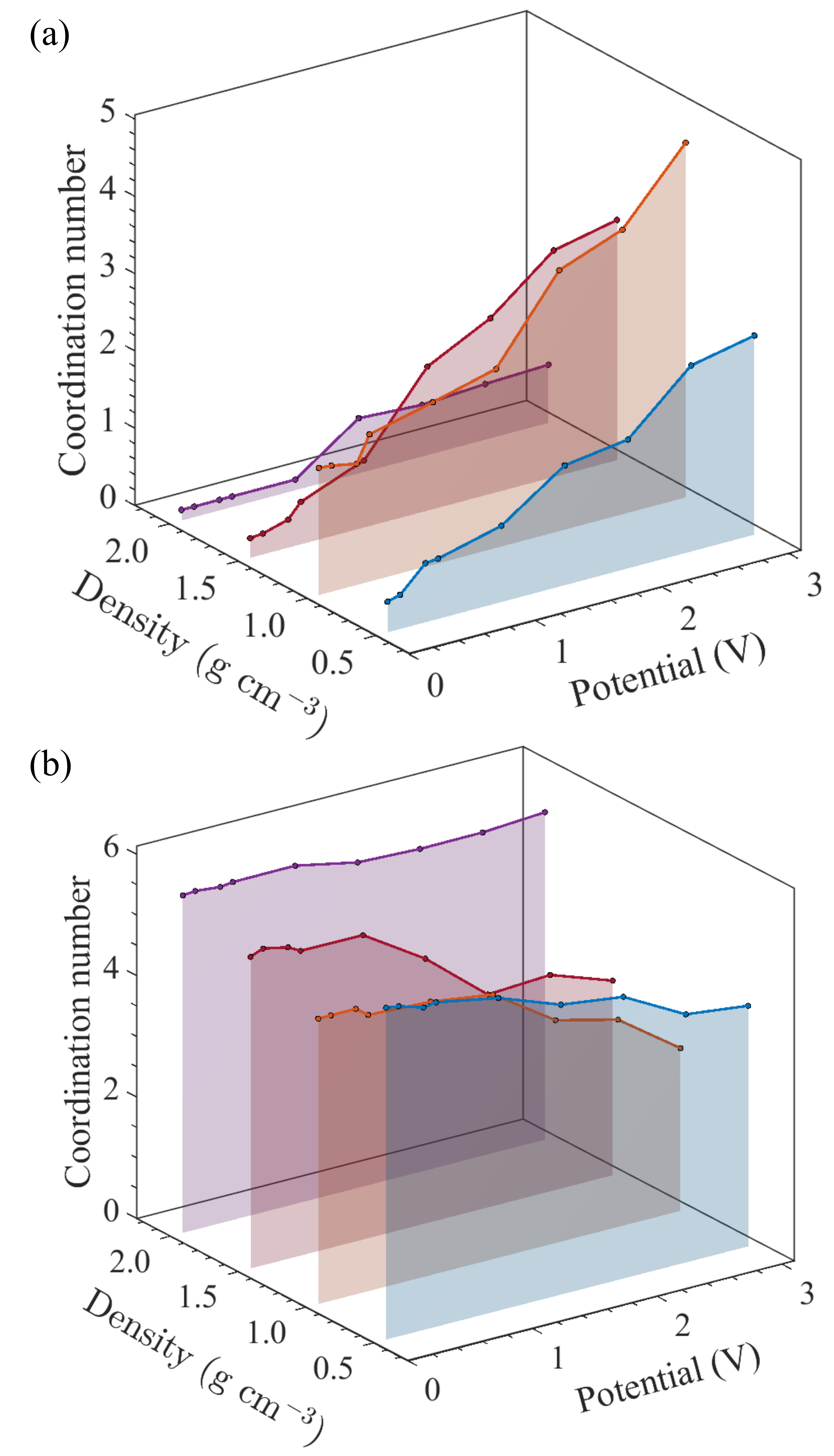}
    \caption{Coordination number (CN) as a function of applied potential and the density of the different carbon electrode models. (a) CN between \ce{Na+} and the carbon electrodes, Na-C. (b) CN between \ce{Na+} and electrolyte oxygen atoms, Na-O.}
    \label{fig:CNs}
\end{figure}

The local coordination between \ce{Na+} and the carbon electrode increases with applied potential for all carbon models (see Figure~\ref{fig:CNs}(a)), indicating enhanced ion–carbon coordination and confinement within the electrode \cite{Lin2025, Li2022b}. At low potentials, Na--C coordination remains low, consistent with ions residing predominantly in the bulk electrolyte or near pore entrances. As the potential increases, \ce{Na+} ions increasingly interact with the carbon structure, leading to a marked increase in Na--C coordination, in agreement with previous atomistic studies of sodium insertion in hard carbon \cite{Lin2025, Li2022, Li2022b}.
The magnitude of this effect is strongly dependent on the electrode density, and therefore porosity. Lower-density carbon electrodes (HC-C1 and HC-C2) exhibit larger and more accessible pore networks, which allow \ce{Na+} to enter a broader range of carbon environments and progressively interact with the pore walls. By contrast, the denser and partially graphitised HC-C3 and HC-C4 nanostructures accommodate considerably fewer additional \ce{Na+} ions. Their coordination behaviour consequently reflects a smaller population of ions sampling more spatially restricted carbon environments. This behaviour is consistent with recent studies demonstrating that pore size and carbon microstructure govern sodium storage mechanisms and ion distribution within hard carbon electrodes~\cite{KitsuIglesias2025, Li2022}.

\begin{figure*}[tb]
    \centering
    \includegraphics[width=0.8\textwidth]{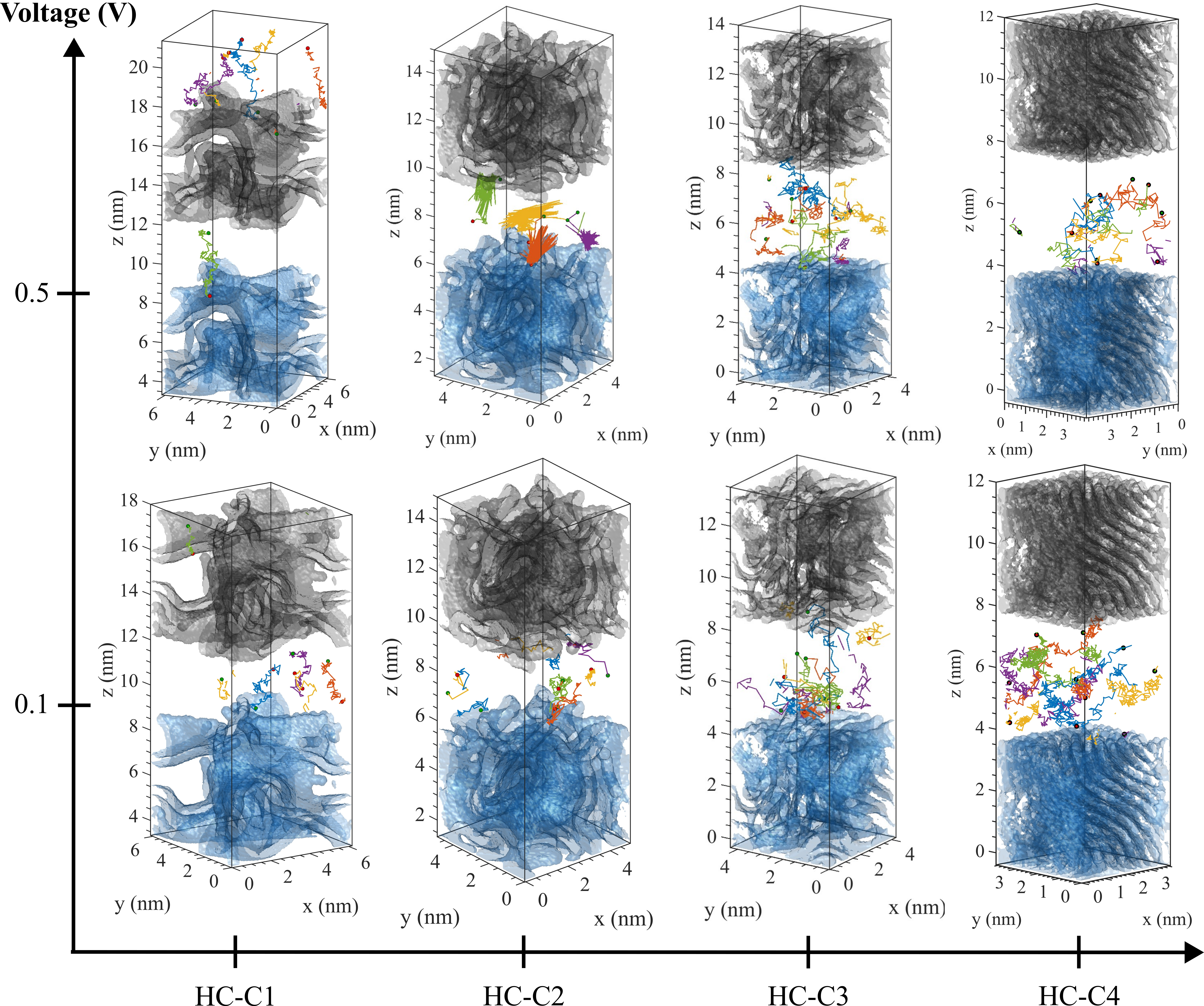}
    \caption{Representative \ce{Na+} trajectories exhibiting the largest net displacement along the electrode normal ($z$) within the HC-C1, HC-C2, HC-C3 and HC-C4 electrode structures at 0.1 and 0.5 V. Coloured lines represent individual \ce{Na+} trajectories, while green and red markers indicate their initial and final positions, respectively. Blue and grey shaded regions represent the opposing porous carbon electrodes.}
    \label{fig:trajectories}
\end{figure*}

The evolution of Na--electrolyte interactions is analysed through the Na--O coordination number, which characterises the local solvation environment of sodium ions (Figure~\ref{fig:CNs}(b)). In contrast to Na--C coordination, Na--O coordination shows a comparatively weak dependence on the applied potential within each carbon model, with values typically between 3 and 5 depending on the carbon density. These values are generally lower than those reported for fully solvated \ce{Na+} ions in bulk carbonate electrolytes, where the first solvation shell typically involves approximately 4--6 oxygen atoms depending on the solvent environment \cite{Shakourian-Fard2015}. This indicates that \ce{Na+} ions sampled in the electrode environment retain substantial, although reduced relative to bulk, solvent coordination under applied potential.

A modest decrease in Na--O coordination is observed at higher potentials for HC-C1, HC-C2 and HC-C3. This trend is consistent with a reduction in the local solvation of \ce{Na+} ions as ions approach and penetrate the carbon electrode, and may therefore reflect partial desolvation at the carbon-electrolyte interface. The observed behaviour is nevertheless consistent with previous studies showing that changes in sodium solvation and desolvation at electrode interfaces can influence sodium-ion storage  and interfacial transport \cite{Goloviznina2023, Liu2023}.

\subsubsection*{Na-ion pathways and electrolyte infiltration revealed by voxelization}

\begin{figure*}[tb]
    \centering
    \includegraphics[width=0.98\textwidth]{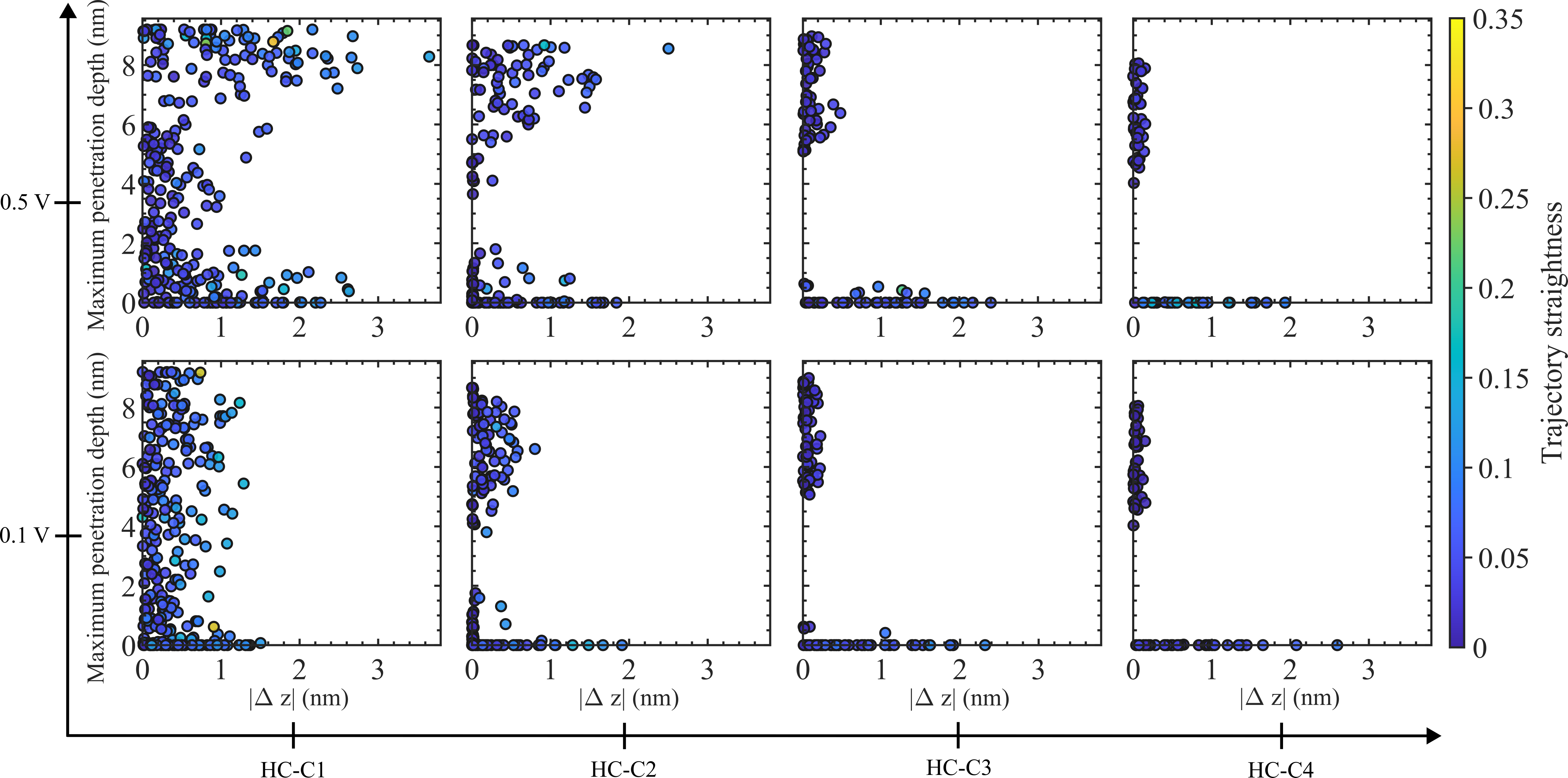}
    \caption{Relationship between sustained \ce{Na+} displacement along the electrode normal, ($|\Delta z|$), and maximum penetration depth for the HC-C1–HC-C4 carbon electrodes at 0.1 and 0.5 V. Each point represents an individual \ce{Na+} trajectory and is coloured according to its trajectory straightness, defined as the ratio between the net three-dimensional displacement and the cumulative trajectory length. Maximum penetration depth represents the deepest position reached by each ion relative to the electrolyte-facing electrode surface, whereas ($|\Delta z|$) quantifies the displacement retained along the electrode normal at the end of the trajectory.}
    \label{fig:migration}
\end{figure*}

\begin{figure*}[tb]
    \centering
    \includegraphics[width=0.98\textwidth]{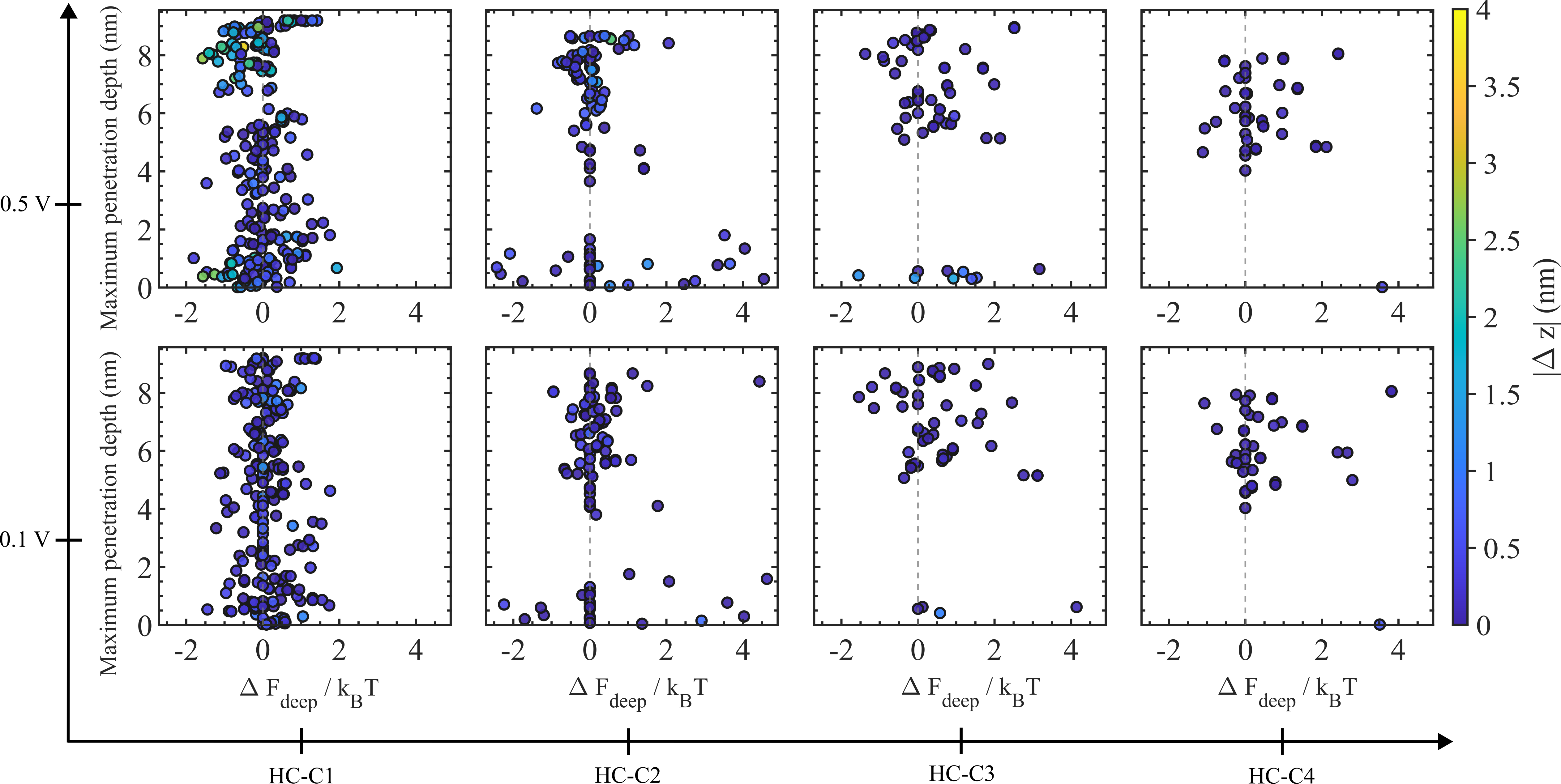}
    \caption{Relationship between maximum \ce{Na+} penetration depth and the PMF-derived free-energy change at the position of deepest penetration, ($\Delta F^{\mathrm{deep}}/k_BT$), for the HC-C1–HC-C4 carbon electrodes at 0.1 and 0.5 V. Each point represents an individual penetrating \ce{Na+} trajectory and is coloured according to its sustained net displacement along the electrode normal, ($|\Delta z|$). The vertical dashed line marks ($\Delta F^{\mathrm{deep}}/k_BT=0$); negative values indicate that the deepest sampled position is more highly populated in the projected PMF than the ion's initial position, whereas positive values indicate a comparatively less populated region.}
    \label{fig:penetration}
\end{figure*}

The coordination analysis indicates that increasing the applied potential promotes stronger interaction of Na$^+$ with the carbon framework, while the comparatively weak variation in Na–O coordination suggests that the ions retain a partially solvated environment as they approach and enter the electrode. However, the coordination numbers alone do not distinguish between transient interfacial interactions and sustained migration through the porous network. To resolve how the underlying carbon topology controls Na$^+$ transport, the trajectories were therefore analysed in terms of their spatial pathways, maximum penetration depth and sustained displacement along the electrode normal. This analysis provides a direct connection between the local coordination environment and the accessibility of the heterogeneous carbon network, allowing us to distinguish ions that merely explore the electrode surface or transiently penetrate the carbon from those that undergo sustained redistribution within the electrode.

Representative Na$^+$ trajectories are shown in Figure~\ref{fig:trajectories} for ions exhibiting the largest net displacement along the electrode normal ($z$). The trajectories reveal strongly heterogeneous transport, with both relatively direct and highly tortuous pathways depending on the underlying carbon microstructure. In HC-C1, the highly porous and strongly curved carbon network provides comparatively open and accessible pathways, resulting in trajectories with a greater degree of directional motion into the electrode interior. As the carbon density increases from HC-C1 to HC-C4, the progressive reduction in pore size and development of more graphitic, planar carbon domains increasingly constrain ion motion, leading to more tortuous trajectories and more pronounced lateral motion near the electrode surface. At 0.5 V, this behaviour is particularly evident in HC-C2, where several Na$^+$ ions exhibit pronounced lateral and back-and-forth motion near the carbon–electrolyte interface. This contrasts with the comparatively more direct trajectories observed in the micro-mesoporous HC-C1 and suggests that, in the microporous HC-C2 structure, the applied potential promotes lateral exploration of available pore entrances and interfacial pathways, with ions undergoing less direct inward migration into the carbon network. The increasingly constrained pore environments (from HC-C2 to HC-C4) therefore limit sustained inward migration, despite the electrostatic driving force.

Figure~\ref{fig:penetration} quantifies this behaviour by comparing the maximum penetration depth with the sustained displacement along the electrode normal, $|\Delta z|$. HC-C1 exhibits the broadest distribution of both quantities, with ions spanning negligible to deep penetration and a comparatively wide range of net displacements. In HC-C2–HC-C4, the distributions become increasingly separated between ions remaining close to the electrode–electrolyte interface and ions reaching deeper regions of the electrode, with comparatively fewer intermediate penetration depths. Importantly, many of the ions reaching deep regions of HC-C2–HC-C4 retain only a small $|\Delta z|$, indicating that their deep penetration is not accompanied by substantial retained displacement along the electrode normal. Maximum penetration and net displacement therefore describe distinct aspects of Na$^+$ transport: an ion can transiently explore a deep region of the electrode without undergoing sustained redistribution into that region. The generally low trajectory straightness across all four structures further indicates that transport through the carbon is predominantly tortuous, with HC-C1 exhibiting the broadest range of trajectory geometries.

Figure~\ref{fig:penetration} relates maximum penetration depth to the PMF-derived free-energy change at the deepest position sampled by each penetrating Na$^+$ ion, $\Delta F_{\mathrm{deep}}/k_{\mathrm B}T$. HC-C1 spans a broad range of penetration depths and free-energy changes, with deeply penetrating ions occurring at both negative and positive $\Delta F_{\mathrm{deep}}/k_{\mathrm B}T$. The other carbon structures similarly sample both sides of $\Delta F_{\mathrm{deep}}/k_{\mathrm B}T=0$, indicating that deep penetration is not associated with a single projected free-energy change. Moreover, HC-C1 contains ions that combine deep penetration with appreciable sustained displacement, whereas deeply penetrating ions in HC-C2–HC-C4 predominantly retain small net displacements. Together, the trajectory and PMF analyses show that reaching a deep position within the carbon does not necessarily correspond to sustained Na$^+$ migration into that region. Instead, the persistence and geometry of Na$^+$ transport emerge from the interplay between the electrostatic driving force and the heterogeneous accessibility and confinement imposed by the carbon microstructure.

\section{Conclusion}

We have investigated the relationship between realistic hard-carbon microstructure and Na$^+$ storage and transport using molecular dynamics simulations of machine-learning-generated carbon structures in an explicit NaPF$_6$/EC:DMC electrolyte. By systematically varying the carbon density from 0.5 to 2.0 g cm$^{-3}$, the simulations provide a controlled structural series spanning highly porous, strongly curved carbon networks to dense structures containing increasingly graphitic local environments.

The electrochemical response demonstrates that electrode charging and Na$^+$ uptake are coupled but non-equivalent processes. Increasing the applied potential increases both electrode polarisation and Na–C coordination, but the extent of Na$^+$ uptake depends strongly on the accessibility and connectivity of the carbon network. In particular, HC-C4 can sustain an increasing electrostatic charge response while exhibiting only very limited additional Na$^+$ uptake. Thus, electrostatic driving force alone is insufficient to produce substantial ionic storage when the underlying carbon structure does not provide accessible transport pathways.

The trajectory analysis provides a more detailed picture of this structural constraint. The open HC-C1 network supports a broader range of penetration depths and sustained Na$^+$ displacements, whereas increasing carbon density progressively confines ion motion and produces predominantly tortuous pathways. Importantly, maximum penetration depth and sustained displacement describe different aspects of transport. Na$^+$ ions in the denser structures can reach relatively deep regions of the carbon while retaining little net displacement, indicating transient penetration followed by redistribution towards the interface rather than sustained migration into the electrode. The applied potential can modify these pathways, as illustrated particularly by the lateral and back-and-forth motion observed in HC-C2, but the structural accessibility of the carbon network remains a fundamental constraint.

The PMF analysis further shows that deep penetration cannot be associated with a single projected free-energy change. Deeply penetrating ions occur across both positive and negative values of $\Delta F_{\mathrm{deep}}/k_{\mathrm B}T$, while the combination of deep penetration and substantial sustained displacement is more prevalent in the open HC-C1 structure. This demonstrates that Na$^+$ transport cannot be characterised solely by whether an ion reaches a particular depth or by the projected free-energy landscape along the electrode normal. Instead, the persistence of migration reflects the combined effects of electrostatic driving force, electrolyte solvation, pore accessibility and the three-dimensional topology of the carbon network.

Overall, these results show that realistic carbon microstructure plays a central role in determining whether electrode polarisation translates into Na$^+$ penetration and sustained redistribution. The distinction between transient pore exploration and persistent ionic transport provides a useful framework for interpreting sodium storage in structurally heterogeneous carbons and highlights the importance of designing accessible, interconnected carbon networks rather than considering electrode polarisation alone.

\begin{acknowledgement}
C.d.T. acknowledges funding from the European Union’s Horizon
2020 research and innovation programme under the Marie Sklodowska-Curie grant agreement No. 101025294.
\end{acknowledgement}


   


  





\bibliography{biblio.bib}

@article{Franzblau,
  title = {Computation of ring statistics for network models of solids},
  author = {Franzblau, D. S.},
  journal = {Phys. Rev. B},
  volume = {44},
  number = {10},
  pages = {4925--4930},
  year = {1991},
  month = {Sep},
  doi = {10.1103/PhysRevB.44.4925},
}

@article{DeTomas2017,
title = {Structural prediction of graphitization and porosity in carbide-derived carbons},
author = {Carla {de Tomas} and Irene Suarez-Martinez and Fernando Vallejos-Burgos and María J. López and Katsumi Kaneko and Nigel A. Marks},
journal = {Carbon},
volume = {119},
pages = {1-9},
year = {2017},
doi = {10.1016/j.carbon.2017.04.004},
}

@article{DeTomas2018APL,
author = {de Tomas,Carla  and Suarez-Martinez,Irene  and Marks,Nigel A. },
title = {Carbide-derived carbons for dense and tunable 3D graphene networks},
journal = {Appl. Phys. Lett.},
volume = {112},
number = {25},
pages = {251907},
year = {2018},
doi = {10.1063/1.5030136},
}

@article{deTomas2023,
year = {2023},
month = {mar},
volume = {5},
number = {2},
pages = {024006},
author = {de Tomas, Carla and Alabidun, Sarat and Chater, Luke and Darby, Matthew T and Raffone, Federico and Restuccia, Paolo and Au, Heather and Titirici, Magda M and Cucinotta, Clotilde S and Crespo-Ribadenyra, Maria},
title = {Doping carbon electrodes with sulfur achieves reversible sodium ion storage},
journal = {Journal of Physics: Energy},
doi = {10.1088/2515-7655/acb570},
}

@book{Warren1969,
author={B. E. Warren},
title={{X-Ray diffraction}},
year={1969},
publisher={{Addison-Wesley Publishing Co., Reading, Massachusetts}},
}

@article{DeTomas2016,
title = "Graphitization of amorphous carbons: A comparative study of interatomic potentials",
journal = "Carbon",
volume = "109",
pages = "681 - 693",
year = "2016",
issn = "0008-6223",
doi = "10.1016/j.carbon.2016.08.024",
author = "Carla de Tomas and Irene Suarez-Martinez and Nigel A. Marks"
}

@article{Plimpton1995,
title = "Fast Parallel Algorithms for Short-Range Molecular Dynamics ",
journal = "J.  Chem. Phys. ",
volume = "117",
number = "1",
pages = "1--19",
year = "1995",
issn = "0021-9991",
doi = "10.1006/jcph.1995.1039",
author = "Plimpton, S"
}

@article{Bartok2013,
  title = {On representing chemical environments},
  author = {Bart\'ok, Albert P. and Kondor, Risi and Cs\'anyi, G\'abor},
  journal = {Phys. Rev. B},
  volume = {87},
  issue = {18},
  pages = {184115},
  numpages = {16},
  year = {2013},
  month = {May},
  publisher = {American Physical Society},
  doi = {10.1103/PhysRevB.87.184115},
}

@article{Padua2017,
author = {P{\'{a}}dua, Ag{\'{i}}lio A. H.},
doi = {10.1063/1.4983687},
issn = {0021-9606},
journal = {The Journal of Chemical Physics},
year = {2017},
volume = {146},
number = {20},
pages  = {204501},
title = {{Resolving dispersion and induction components for polarisable molecular simulations of ionic liquids}},
url = {https://github.com/paduagroup/fftool https://pubs.aip.org/jcp/article/146/20/204501/671642/Resolving-dispersion-and-induction-components-for},
}

@article{Martinez2009,
author = {Mart{\'{i}}nez, L. and Andrade, R. and Birgin, E. G. and Mart{\'{i}}nez, J. M.},
doi = {10.1002/jcc.21224},
issn = {0192-8651},
journal = {Journal of Computational Chemistry},
month = {oct},
number = {13},
pages = {2157--2164},
title = {{PACKMOL: A package for building initial configurations for molecular dynamics simulations}},
url = {https://onlinelibrary.wiley.com/doi/10.1002/jcc.21224},
volume = {30},
year = {2009}
}

@article{Wu2024,
   author = {Chun Wu and Yunrui Yang and Yinghao Zhang and Hui Xu and Xiangxi He and Xingqiao Wu and Shulei Chou},
   doi = {10.1039/d4sc00734d},
   issn = {20416539},
   issue = {17},
   journal = {Chemical Science},
   month = {3},
   pages = {6244-6268},
   publisher = {Royal Society of Chemistry},
   title = {Hard carbon for sodium-ion batteries: progress, strategies and future perspective},
   volume = {15},
   year = {2024}
}

@article{KitsuIglesias2025,
   author = {Luis Kitsu Iglesias and Samuel D. Marks and Nikhil Rampal and Emma N. Antonio and Rafael de Ferreira de Menezes and Liang Zhang and Daniel Olds and Stephen E. Weitzner and Kayla G. Sprenger and Liwen F. Wan and Michael F. Toney},
   doi = {10.1002/smll.202505561},
   issn = {16136829},
   issue = {30},
   journal = {Small},
   month = {7},
   pmid = {40442960},
   publisher = {John Wiley and Sons Inc},
   title = {Microstructure-Dependent Sodium Storage Mechanisms in Hard Carbon Anodes},
   volume = {21},
   year = {2025}
}

@article{Lin2025,
   author = {Che an Lin and Huu Duc Luong and Ryoma Sasaki and Yoshitaka Tateyama},
   doi = {10.1002/aenm.202505227},
   issn = {16146840},
   journal = {Advanced Energy Materials},
   month = {1},
   publisher = {John Wiley and Sons Inc},
   title = {Unveiling Dominant Processes of Na Cluster Formation and Na-Ion Diffusion in Hard Carbon Nano-Pore: A DFT-MD Study},
   year = {2025}
}

@article{Li2022,
   author = {Jiaqi Li and Chen Peng and Jingkun Wang and Jie Li and Hongliang Zhang},
   doi = {10.1016/j.diamond.2022.109355},
   issn = {09259635},
   journal = {Diamond and Related Materials},
   month = {11},
   publisher = {Elsevier Ltd},
   title = {Towards an atomistic understanding of hard carbon electrode materials and sodium behaviors},
   volume = {129},
   year = {2022}
}

@article{Hjertens2016,
   author = {Eirik Hjertenæs and Anh Quynh Nguyen and Henrik Koch},
   doi = {10.1039/c6cp06774c},
   issn = {14639076},
   issue = {46},
   journal = {Physical Chemistry Chemical Physics},
   pages = {31431-31440},
   publisher = {Royal Society of Chemistry},
   title = {A ReaxFF force field for sodium intrusion in graphitic cathodes},
   volume = {18},
   year = {2016}
}

@article{Li2022a,
   author = {Jiaqi Li and Jie Li and Jingkun Wang and Hongliang Zhang},
   doi = {10.1016/j.cplett.2022.139657},
   issn = {00092614},
   journal = {Chemical Physics Letters},
   month = {8},
   publisher = {Elsevier B.V.},
   title = {Insight into sodium penetration with mechanical behaviors of carbon electrodes by large-scale reactive molecular dynamics simulations},
   volume = {800},
   year = {2022}
}

@article{Li2022b,
   author = {Jiaqi Li and Chen Peng and Jie Li and Jingkun Wang and Hongliang Zhang},
   doi = {10.1021/acs.energyfuels.2c00575},
   issn = {15205029},
   issue = {11},
   journal = {Energy and Fuels},
   month = {6},
   pages = {5937-5952},
   publisher = {American Chemical Society},
   title = {Insight into Sodium Storage Behaviors in Hard Carbon by ReaxFF Molecular Dynamics Simulation},
   volume = {36},
   year = {2022}
}

@article{Ibragimova2025,
   author = {Rina Ibragimova and Mikhail S. Kuklin and Tigany Zarrouk and Miguel A. Caro},
   doi = {10.1021/acs.chemmater.4c02905},
   issn = {15205002},
   issue = {3},
   journal = {Chemistry of Materials},
   month = {2},
   pages = {1094-1110},
   publisher = {American Chemical Society},
   title = {Unifying the Description of Hydrocarbons and Hydrogenated Carbon Materials with a Chemically Reactive Machine Learning Interatomic Potential},
   volume = {37},
   year = {2025}
}

@article{Zarrouk2024,
   author = {Tigany Zarrouk and Rina Ibragimova and Albert P. Bartók and Miguel A. Caro},
   doi = {10.1021/jacs.4c01897},
   issn = {15205126},
   issue = {21},
   journal = {Journal of the American Chemical Society},
   month = {5},
   pages = {14645-14659},
   pmid = {38749497},
   publisher = {American Chemical Society},
   title = {Experiment-Driven Atomistic Materials Modeling: A Case Study Combining X-Ray Photoelectron Spectroscopy and Machine Learning Potentials to Infer the Structure of Oxygen-Rich Amorphous Carbon},
   volume = {146},
   year = {2024}
}

@article{Liu2026,
   author = {Xinyu Liu and Kara Fong and Zhaohan Shen and Wei Yu and Hirotomo Nishihara and Clare P. Grey and Alexander C. Forse},
   doi = {10.1021/acselectrochem.5c00472},
   issn = {2997-0571},
   journal = {ACS Electrochemistry},
   month = {2},
   publisher = {American Chemical Society (ACS)},
   title = {Electrolyte Effects on Disorder-Enhanced Capacitance in Nanoporous Carbons},
   year = {2026}
}

@article{Liu2023,
   author = {Guiyu Liu and Zhiqiang Wang and Huimin Yuan and Chunliu Yan and Rui Hao and Fangchang Zhang and Wen Luo and Hongzhi Wang and Yulin Cao and Shuai Gu and Chun Zeng and Yingzhi Li and Zhenyu Wang and Ning Qin and Guangfu Luo and Zhouguang Lu},
   doi = {10.1002/advs.202305414},
   issn = {21983844},
   issue = {36},
   journal = {Advanced Science},
   month = {12},
   pmid = {37875394},
   publisher = {John Wiley and Sons Inc},
   title = {Deciphering Electrolyte Dominated Na+ Storage Mechanisms in Hard Carbon Anodes for Sodium-Ion Batteries},
   volume = {10},
   year = {2023}
}

@article{Goloviznina2023,
   author = {Kateryna Goloviznina and Ezzoubair Bendadesse and Ozlem Sel and Jean Marie Tarascon and Mathieu Salanne},
   doi = {10.1021/acsami.3c12815},
   issn = {19448252},
   issue = {51},
   journal = {ACS Applied Materials and Interfaces},
   month = {12},
   pages = {59380-59388},
   pmid = {38095112},
   publisher = {American Chemical Society},
   title = {Disclosing the Interfacial Electrolyte Structure of Na-Insertion Electrode Materials: Origins of the Desolvation Phenomenon},
   volume = {15},
   year = {2023}
}

@article{Shakourian-Fard2015,
   author = {Mehdi Shakourian-Fard and Ganesh Kamath and Kassiopeia Smith and Hui Xiong and Subramanian K.R.S. Sankaranarayanan},
   doi = {10.1021/acs.jpcc.5b04706},
   issn = {19327455},
   issue = {40},
   journal = {Journal of Physical Chemistry C},
   month = {10},
   pages = {22747-22759},
   publisher = {American Chemical Society},
   title = {Trends in Na-Ion Solvation with Alkyl-Carbonate Electrolytes for Sodium-Ion Batteries: Insights from First-Principles Calculations},
   volume = {119},
   year = {2015}
}

@article{yabuuchi2014research,
  title={Research development on sodium-ion batteries},
  author={Yabuuchi, Naoaki and Kubota, Kei and Dahbi, Mouad and Komaba, Shinichi},
  journal={Chemical Reviews},
  volume={114},
  number={23},
  pages={11636--11682},
  year={2014},
  publisher={ACS Publications},
  doi={10.1021/cr500192f},
}

@article{hwang2017sodium,
  title={Sodium-ion batteries: present and future},
  author={Hwang, Jang-Yeon and Myung, Seung-Taek and Sun, Yang-Kook},
  journal={Chemical Society Reviews},
  volume={46},
  number={12},
  pages={3529--3614},
  year={2017},
  publisher={The Royal Society of Chemistry},
  doi={10.1039/c6cs00776g},
}

@article{cai2021comprehensive,
  title={Comprehensive understanding of sodium-ion capacitors: definition, mechanisms, configurations, materials, key technologies, and future developments},
  author={Cai, Peng and Zou, Kangyu and Deng, Xinglan and Wang, Baowei and Zheng, Min and Li, Longhao and Hou, Hongshuai and Zou, Guoqiang and Ji, Xiaobo},
  journal={Advanced Energy Materials},
  volume={11},
  number={16},
  pages={2003804},
  year={2021},
  publisher={Wiley Online Library},
  doi={10.1002/aenm.202003804},
}

@article{vaalma2018cost,
  title={A cost and resource analysis of sodium-ion batteries},
  author={Vaalma, Christoph and Buchholz, Daniel and Weil, Marcel and Passerini, Stefano},
  journal={Nature reviews materials},
  volume={3},
  number={4},
  pages={18013},
  year={2018},
  publisher={Nature Publishing Group},
  doi={ 10.1038/natrevmats.2018.13},
}

@article{Stevens_2000,
doi = {10.1149/1.1393348},
year = {2000},
month = {apr},
publisher = {The Electrochemical Society, Inc.},
volume = {147},
number = {4},
pages = {1271},
author = {Stevens, D. A. and Dahn, J. R.},
title = {High Capacity Anode Materials for Rechargeable Sodium‐Ion Batteries},
journal = {Journal of The Electrochemical Society}
}

@article{bommier2015new,
  title={New mechanistic insights on Na-ion storage in nongraphitizable carbon},
  author={Bommier, Clement and Surta, Todd Wesley and Dolgos, Michelle and Ji, Xiulei},
  journal={Nano Letters},
  volume={15},
  number={9},
  pages={5888--5892},
  year={2015},
  publisher={ACS Publications},
  doi={10.1021/acs.nanolett.5b01969},
}

@article{Muck2025,
author = {Mück, Stefan and Kramer, Dominik and Palanisamy, Krishnaveni and Kranz, Christine and Mönig, Reiner},
title = {Mechanical Stress Reveals Asymmetry of Sodiation and Desodiation of Hard Carbon},
journal = {ChemSusChem},
volume = {18},
number = {19},
pages = {e202501272},
doi = {10.1002/cssc.202501272},
url = {https://chemistry-europe.onlinelibrary.wiley.com/doi/abs/10.1002/cssc.202501272},
eprint = {https://chemistry-europe.onlinelibrary.wiley.com/doi/pdf/10.1002/cssc.202501272},
year = {2025}
}

@article{Olsson2025,
author = {Olsson, Martina and Klein, Antoine and Mozhzhukhina, Nataliia and Xiong, Shizhao and Appel, Christian and Carlsen, Mads and Nielsen, Leonard and Rensmo, Linnea and Liebi, Marianne and Matic, Aleksandar},
title = {Resolving the Sodiation Process in Hard Carbon Anodes with Nanostructure Specific X-Ray Imaging},
journal = {Advanced Science},
volume = {12},
number = {34},
pages = {e08635},
doi = {10.1002/advs.202508635},
url = {https://advanced.onlinelibrary.wiley.com/doi/abs/10.1002/advs.202508635},
eprint = {https://advanced.onlinelibrary.wiley.com/doi/pdf/10.1002/advs.202508635},
year = {2025}
}

@article{kirkland2016,
  title={Computation in electron microscopy},
  author={Kirkland, Earl J},
  journal={Acta Crystallographica Section A: Foundations and Advances},
  volume={72},
  number={1},
  pages={1--27},
  year={2016},
  publisher={International Union of Crystallography},
  doi = {10.1107/S205327331501757X},

}

@article{Guo-Titirici2023,
author = {Guo, Zhenyu and Xu, Zhen and Xie, Fei and Jiang, Jinglin and Zheng, Kaitian and Alabidun, Sarat and Crespo-Ribadeneyra, Maria and Hu, Yong-Sheng and Au, Heather and Titirici, Maria-Magdalena},
title = {Investigating the Superior Performance of Hard Carbon Anodes in Sodium-Ion Compared With Lithium- and Potassium-Ion Batteries},
journal = {Advanced Materials},
volume = {35},
number = {42},
pages = {2304091},
doi = {10.1002/adma.202304091},
year = {2023}
}

@article{Wang2026,
  author  = {Wang, Z. and Shi, G. and Wang, G. and Wang, M. and Ding, F. and Wang, X.},
  title   = {Unveiling sodium storage mechanisms in hard carbon via machine learning-driven simulations integrated with accurate site occupation identification},
  journal = {Chemical Science},
  volume  = {17},
  number  = {5},
  pages   = {2547--2558},
  year    = {2026},
  doi     = {10.1039/d5sc07068f},
}

@article{Sarkisov2020,
    author = {Sarkisov, Lev and Bueno-Perez, Rocio and Sutharson, Mythili and Fairen-Jimenez, David},
    title = {Materials Informatics with PoreBlazer v4.0 and the
CSD MOF Database},
    journal = {Chemistry of Materials},
    volume = {32},
    number = {23},
    pages = {9849-9867},
    year = {2020},
    month = {11},
    issn = {0897-4756},
    doi = {10.1021/acs.chemmater.0c03575},
   
}

@article{Front2026,
title = {Density-dependent sodium-storage mechanisms in hard carbon materials},
journal = {Chemical Science},
volume = {17},
number = {19},
pages = {9570-9581},
year = {2026},
issn = {2041-6520},
doi = {10.1039/d6sc00030d},
author = {Alexis Front and Tapio Ala-Nissilä and Miguel A. Caro}
}

@article{Komaba2011,
author = {Komaba, Shinichi and Murata, Wataru and Ishikawa, Toru and Yabuuchi, Naoaki and Ozeki, Tomoaki and Nakayama, Tetsuri and Ogata, Atsushi and Gotoh, Kazuma and Fujiwara, Kazuya},
title = {Electrochemical Na Insertion and Solid Electrolyte Interphase for Hard-Carbon Electrodes and Application to Na-Ion Batteries},
journal = {Advanced Functional Materials},
volume = {21},
number = {20},
pages = {3859-3867},
doi = {10.1002/adfm.201100854},
year = {2011}
}

@article{Eren2025,
    author = {Eren, Enis Oğuzhan and Senokos, Evgeny and Scoppola, Ernesto and Song, Zihan and Antonietti, Markus and Giusto, Paolo},
    title = {An enhanced three-stage model for sodium storage in hard carbons},
    journal = {Energy \& Environmental Science},
    volume = {18},
    number = {16},
    pages = {7859-7868},
    year = {2025},
    month = {08},
    issn = {1754-5692},
    doi = {10.1039/d4ee06029f},
}

@article{Alptekin2020,
    author = {Alptekin, Hande and Au, Heather and Jensen, Anders CS and Olsson, Emilia and Goktas, Mustafa and Headen, Thomas F. and Adelhelm, Philipp and Cai, Qiong and Drew, Alan J and Titirici, Maria-Magdalena},
    title = {Sodium
Storage Mechanism Investigations through Structural
Changes in Hard Carbons},
    journal = {ACS Applied Energy Materials},
    volume = {3},
    number = {10},
    pages = {9918-9927},
    year = {2020},
    month = {09},
    issn = {2574-0962},
    doi = {10.1021/acsaem.0c01614},
}

@article{Au2020,
    author = {Au, Heather and Alptekin, Hande and Jensen, Anders C. S. and Olsson, Emilia and O’Keefe, Christopher A. and Smith, Thomas and Crespo-Ribadeneyra, Maria and Headen, Thomas F. and Grey, Clare P. and Cai, Qiong and Drew, Alan J. and Titirici, Maria-Magdalena},
    title = {A revised mechanistic model for sodium insertion in hard carbons},
    journal = {Energy \& Environmental Science},
    volume = {13},
    number = {10},
    pages = {3469-3479},
    year = {2020},
    month = {10},
    issn = {1754-5692},
    doi = {10.1039/d0ee01363c},
}

@article{Rowe2020,
  title={An accurate and transferable machine learning potential for carbon},
  author={Rowe, Patrick and Deringer, Volker L and Gasparotto, Piero and Cs{\'a}nyi, G{\'a}bor and Michaelides, Angelos},
  journal={The Journal of Chemical Physics},
  volume={153},
  number={3},
  year={2020},
  publisher={AIP Publishing}
}

@article{AhrensIwers2022,
  title={ELECTRODE: An electrochemistry package for atomistic simulations},
  author={Ahrens-Iwers, Ludwig JV and Janssen, Mathijs and Tee, Shern R and Mei{\ss}ner, Robert H},
  journal={The journal of chemical physics},
  volume={157},
  number={8},
  year={2022},
  publisher={AIP Publishing}
}

@article{Nanoporous-carbonsTurbogap,
  title={Structure and pore size distribution in nanoporous carbon},
  author={Wang, Yanzhou and Fan, Zheyong and Qian, Ping and Ala-Nissila, Tapio and Caro, Miguel A},
  journal={Chemistry of Materials},
  volume={34},
  number={2},
  pages={617--628},
  year={2022},
  publisher={ACS Publications}
}

\end{document}